\documentclass[notoc,12pt]{article}
\usepackage{amsfonts}
\usepackage{amscd}
\usepackage{amssymb}
\usepackage{amsmath}
\usepackage{epsfig}
\usepackage{latexsym}
\usepackage{appendix}

\def \be  {\begin{equation}}
\def \ee  {\end{equation}}
\def \ba  {\begin{eqnarray}}
\def \ea  {\end{eqnarray}}
\def \bb  {\begin{thebibliography}}
\def \eb  {\end{thebibliography}}
\def \lab #1 {\label{#1}}
\newcommand\M{\mathbb{M}}

\newcommand\cA{\mathcal{A}}

\newcommand\C {\mathbb{C }}
\newcommand\R {\mathbb{R }}

\newcommand\CP {\mathbb{CP}}

\newcommand\rd{\mathrm{\, d}}

\newcommand\Li{\mathrm{Li}}

\newcommand\dbar{\bar\partial}
\newcommand\la{\langle}
\newcommand\ra{\rangle}

\newcommand\tr{\mathrm{Tr}}

\begin{document}

\title{From the Holomorphic Wilson Loop to `d log' Loop-Integrands of Super-Yang-Mills Amplitudes}
\author{\normalsize Arthur E. Lipstein \& Lionel Mason \\ \small \textit{The Mathematical Institute} \\ \small \textit{University of Oxford} \\ \small \textit{24-29 St Giles'} \\ \small \textit{Oxford, OX1 3LB, U.K.}}
\maketitle
\begin{abstract}
The S-matrix for  planar $\mathcal{N}=4$ super Yang-Mills theory can be computed as the correlation function for a holomorphic polygonal Wilson loop in twistor space.  In an axial gauge, this leads to the construction of the all-loop integrand via MHV diagrams in  twistor space.  We show that at MHV, this formulation leads directly to expressions for loop integrands in d log form; i.e., the integrand is a product of exterior derivatives of logarithms of rational functions.  For higher MHV degree, it is in d log form multiplied by delta functions. The parameters appearing in the d log form arise geometrically as the coordinates of insertion points of propagators on the holomorphic Wilson loop or on MHV vertices.   We discuss a number of examples at one and two loops and give a preliminary discussion of the  evaluation of the 1-loop MHV amplitude. 
{\normalsize \par}
\end{abstract}

\section{Introduction}
Many powerful techniques have been developed to compute scattering amplitudes of 4d $\mathcal{N}=4$ super
Yang-Mills (SYM) \cite{Brink:1976bc} following on from Witten's discovery of twistor-string theory \cite{Witten:2003nn}.  SYM amplitudes can be computed
using BCFW recursion relations (which relate higher point on-shell
amplitudes to lower-point on-shell amplitudes)
\cite{Britto:2004ap,Britto:2005fq,ArkaniHamed:2010kv} or an MHV
formalism (where Maximal Helicity Violating, MHV, amplitudes are used
as the Feynman vertices for constructing all other amplitudes)
\cite{Cachazo:2004kj,Risager:2005vk}. 
In a separate development, planar on-shell scattering amplitudes of $\mathcal{N}=4$ sYM were conjectured to be
dual to null polygonal Wilson loops
\cite{Alday:2007hr,Brandhuber:2007yx,Drummond:2007cf,
Dualc,Drummond:2007aua,Brandhuber:2008pf,Drummond:2008vq}.  The superconformal symmetry of the Wilson loop is distinct from that for the amplitude and has become known as dual superconformal symmetry.  When it is combined with ordinary superconformal symmetry, this gives Yangian symmetry \cite{Dolan:2004ps,Drummond:2009fd}.  Recently,
a Wilson loop/correlator duality has also been discovered
\cite{Alday:2010zy,Eden:2010zz,Eden:2010ce,Eden:2011yp,Eden:2011ku,Adamo:2011dq}.

Many further insights emerged when these ideas were realized in twistor space.  Twistors had for a long time been used as a tool to study conformal invariance, and were introduced in the context of dual conformal symmetry as momentum twistors in \cite{Hodges:2009hk}.
The MHV expansion of $\mathcal{N}=4$ sYM can be obtained as
the Feynman diagrams of a
twistor space action in an axial gauge both for amplitudes
\cite{Boels:2006ir,Boels:2007qn,Adamo:2011cb} (where the
usual superconformal symmetry is manifest) or for the null polygonal
Wilson loop re-expressed as a holomorphic Wilson-loop in twistor space \cite{Mason:2010yk,Bullimore:2010pj} where dual conformal symmetry is manifest. The twistor action can also be used to obtain an analytic proof at the level of the loop integrand of the supersymmetric amplitude/Wilson loop duality \cite{Mason:2010yk,Bullimore:2011ni} (see also \cite{CaronHuot:2010ek} for a space-time formulation). 

A crucial construction in this approach is the holomorphic Wilson loop in twistor space.   Classically it is a reformulation in twistor space of the null polygonal Wilson loop in space-time, but in twistor space it is more canonically defined, being defined manifestly supersymmetrically and  off-shell.  Its space-time counterpart cannot be realised manifestly super-symmetrically and can only be canonically defined on-shell leading to ambiguities and difficulties in the quantum correlation function that have not yet been fully resolved \cite{Belitsky:2011zm}, although progress is now being made \cite{Belitsky:2012rc}.    The holomorphic Wilson loop has a simple Feynman expansion in an axial gauge that corresponds directly to the MHV diagram expansion for amplitudes.  Thus the MHV formalism manifests both ordinary superconformal invariance \cite{Adamo:2011cb} and dual superconformal invariance \cite{Bullimore:2010pj} but for the choice of the`reference twistor' used to construct the axial gauge.  Hence, in this framework, the MHV formalism is an off shell Feynman diagram formalism that manifests  the full Yangian symmetry up to a the choice of a reference twistor.  The compactness of the formalism allows the Risager tree-level proof of the MHV rules to be extended to the full planar loop integrand \cite{Bullimore:2010dz}.
The Q-bar anomaly equations satisfied by amplitudes and null polygonal Wilson loops \cite{CaronHuot:2011kk}  is naturally interpreted in this formulation \cite{Bullimore:2011kg}.   The off-shell framework allows us to study  correlation functions of other gauge invariant operators also, and indeed leads to a manifestly supersymmetric formulation and proof of the correlation function/Wilson loop correspondence \cite{Adamo:2011dq} extending and proving conjectures of previous authors.  
One of the purposes of this paper is to develop further this theory of the holomorphic Wilson-loop in twistor space.

In a separate development, BCFW recursion can be also realised in twistor space \cite{Mason:2009sa,ArkaniHamed:2009si} expressing full superconformal invariance.  This  led to a formula that generates the amplitudes and leading singularities of $\mathcal{N}=4$ sYM via a contour integral over a Grassmannian \cite{ArkaniHamed:2009dn}.  There is a parallel Grassmannian dual conformal invariant formula for the Wilson loop \cite{Mason:2009qx}, and the translation between the two expresses the Yangian symmetry  in this framework \cite{ArkaniHamed:2009vw}.  BCFW recursion was extended to generate the loop integrand in \cite{ArkaniHamed:2010kv,Boels:2010nw}.  For a review of all these twistor developments see \cite{Adamo:2011pv}.

A key long term goal is to actually calculate the planar $n$ particle S-matrix/Wilson-loop explicitly to all loop orders. Although dual superconformal symmetry of the amplitudes is broken by infrared divergences, a solution to the anomalous Ward identity for maximally-helicity violating (MHV) amplitudes
is to write them in terms of the BDS ansatz \cite{Bern:2005iz},
\[
A_{n}^{MHV}=A_{n}^{BDS}\exp R_{n}
\]
where $R_n$ is the so-called remainder function.  It is finite, dual-conformal invariant and nontrivial for $n>5$.  Much work has now gone into the study and evaluation of these remainder functions from a variety of perspectives  \cite{Goncharov:2010jf,Drummond:2007bm,Bern:2008ap,Drummond:2008aq,DelDuca:2009au,DelDuca:2010zg, CaronHuot:2011ky, Dixon:2011pw,CaronHuot:2011kk,Dixon:2011nj}.
At weak coupling, the MHV amplitudes can be expressed in terms of certain transcendental functions known generically as polylogarithms.   
These are usually expressed as iterated indefinite integrals of the dlog of rational functions, the simplest examples being 
$$
\log x=\int^x_1 \frac{ \rd  s}{s}=\int^x_1\rd\log s \, , \quad \Li_2 x= -\int_0^x \log (1-s)\rd \log s\, , \quad \Li_n x=\int ^x_0\Li_{n-1}\, s \, \rd\log s\, .
$$
In general any $\rd\log f$ is allowed in such an iterated integral so long as $f$ is a rational function.  An empirical observation is that it seems to be the case for $\mathcal{N}=4$ SYM that $L$-loop amplitudes are given by polylogarithms arising from $2L$ such iterated integrals (indeed it is expected that amplitudes have precise transcendentality degree $2L$ for $\mathcal{N}=4$, but that lower transcendentality degrees arise also for less supersymmetric theories).

In this paper, we will describe a new form of hidden simplicity in the loop amplitudes of $\mathcal{N}=4$ super Yang-Mills theory. In particular, we will show that for any number of external legs and any number of loops, the amplitude can be reduced to an integral whose integrand is locally in dlog form. Such observations were first made in the context of loop integrands arising from the BCFW recursion and the Grassmannian framework \footnote{As described in a number of lectures in 2012 by Nima Arkani-Hamed, Jake Bourjailly, Freddy Cachazo, Simon Caron-Huot and Jaroslav Trnka in `Scattering Amplitudes', Trento, `String Math', Bonn, `the Geometry of Scattering Amplitudes', Banff, and `Amplitudes and Periods', IHE, and \cite{ArkaniHamed:2012nw}. This work provided some of the impetus for this paper.} and the main purpose of this note is to show how simply and naturally such dlog integrands arise from the holomorphic Wilson loop in twistor space diagram by diagram in its Feynman expansion. In this approach, for an $L$-loop amplitude, we will obtain an integrand that is a differential form that is a wedge product of $4L$ dlogs.  No external data of the scattering amplitude appears in the differential form itself, but it is encoded in the integration contours. For example, we will show that the one-loop MHV amplitude consists of diagrams that can all be reduced to the integral of the $4$-form:
\begin{equation}
\int\frac{ds_{0}}{s_{0}}\frac{dt_{0}}{t_{0}}\frac{ds}{s}\frac{dt}{t}\, .
\label{kermi}
\end{equation}
where $s_0,s_1,t_0,t_1$ are parameters arising from the geometry of the Wilson loop.  They are  related to standard region momentum variables by
\be\label{}
s_0=-\frac{x_{0i}^2}{[\xi|x_{0i}|\bar\xi\ra}\, , \quad t_0=-\frac{x_{0j}^2}{[\xi|x_{0j}|\bar\xi\ra}\, , \quad s= -\frac{[\xi|x_{0i}|\lambda_{i-1}\ra }{[\xi|x_{0i}|\lambda_{i}\ra}\, ,\quad t= -\frac{[\xi|x_{0j}|\lambda_{j-1}\ra }{[\xi|x_{0j}|\lambda_{j}\ra}\, ,
\ee
where $\xi$ is a reference spinor, $(x_0$, $x_i)$ are the loop and external region momenta, and $\lambda_i$ is an external momentum spinor.  The integral is taken over the contour defined by
\begin{equation}
s_{0}=\bar{s}_{0},\,\,\, t_{0}=\bar{t}_{0},
\label{real}
\end{equation}
\begin{equation}
s=-\frac{\bar{t}\left(a_{i-1j}+iv\right)+a_{i-1j-1}+iv}{\bar{t}\left(a_{i\, j}+iv\right)+a_{ij-1}+iv},\,\,\,v=s_0-t_0.
\label{real2}
\end{equation}  
Here $a_{ij}=Z_i \cdot \bar{Z}_j$, where $Z_i$ and $Z_j$ are the momentum twistors of the external particles appearing in the diagram (which we will define later). We will discuss the detailed interpretation of this integral later (it is not strictly speaking well-defined as stated because we need an $i\epsilon$ prescription for the real integrals). This integral  was first studied as the simplest MHV loop diagram in standard momentum space in \cite{Brandhuber:2004yw}.
It was referred to as the `Kermit' diagram in \cite{ArkaniHamed:2010kv,Bullimore:2010pj} where it was expressed in a dual conformally invariant region momentum space form. 

In general, it will be clear that there will be four integrals per loop order in this approach and the integral is over a compact contour. This is in contradistinction to the $2L$ indefinite integrals in the standard definition of polylogs given above.  There is thus a principal that half the integrals have the effect of reducing the other half to indefinite integrals without increasing the transcendentality degree. 

We show that this result generalizes to any number of loops, using the holomorphic Wilson loop in momentum twistor space which is dual to the S-matrix of planar $\mathcal{N}=4$ sYM.  In particular, the dlog form of the integrand follows straightfowardly from the Feynman rules for the twistor space Wilson loop, and the constraints giving rise to the integration contour simply correspond to the reality conditions  on the loop momentum expressed in momentum twistor space. The parameters whose dlogs make up the loop integrand have simple geometrical interpretations on the holomorphic Wilson loop as the location of insertion points of the propagators around the loop or on MHV vertices. Note that the constraints for the Kermit integral in eq \ref{real} and \ref{real2} imply that two of the integration variables are real and the other two are complex. This will hold more generally, as we explain in section \ref{contour}.


This paper is organized as follows. In section \ref{basics}, we briefly review some important concepts like momentum twistors and the holomorphic Wilson loop, and set up the notation for the rest of the paper. In section \ref{wilsonloop}, we give the Feynman rules for the twistor space Wilson loop in a form that naturally leads to dlog forms for the integrands. In section \ref{examples}, we use the formalism in section \ref{wilsonloop} to show how this works in practice for the one and two-loop MHV amplitudes and one the loop NMHV amplitude, and explain how to obtain the contours for these integrals. We present our conclusions in section \ref{conclusion}. In appendix \ref{toy} we study a toy model for the Kermit integral and in appendix \ref{reverse} we show how our conjectured result for the generic contribution to the 1-loop MHV integral (Kermit) gives rise to the symbol of the 1-loop MHV amplitude.        

 \section{Amplitudes and holomorphic Wilson loops} \label{basics}

\subsection{Region momenta and momentum twistor space}
The momentum of a particle in 4d can be written in bispinor form as
follows: \[
p^{\alpha\dot{\alpha}}=\lambda^{\alpha}\tilde{\lambda}^{\dot{\alpha}}\]
where $\alpha=0,1$ and $\dot{\alpha}=\dot{0},\dot{1}$ are chiral
and antichiral spinor indices. In a supersymmetric theory, the particles
also have supermomentum. In particular, the supermomentum of a particle
in $\mathcal{N}=4$ sYM can be written as\[
q^{a\alpha}=\lambda^{\alpha}\eta^{a}\]
where $\eta$ is a fermionic variable and $a$ is an $SU(4)$ R-symmetry
index. Furthermore, an $n$-point superamplitude can be parameterized
in terms of the variables $\left(\lambda_{i}^{\alpha},\tilde{\lambda}_{i}^{\dot{\alpha}},\eta_{i}^{a}\right)$
where $i$ labels the external particles, i.e. $i=1,...,n$. For example,
an $n$-point MHV superamplitude has the following very simple form 

\[
A_{n}^{MHV}=\frac{\delta^{4}\left(p\right)\delta^{8}\left(q\right)}{\left\langle 12\right\rangle \left\langle 23\right\rangle ...\left\langle n1\right\rangle }\]
where $p=\sum_{i=1}^{n}p_{i}$, $q=\sum_{i=1}^{n}q_{i}$, and $\left\langle ij\right\rangle =\epsilon_{\alpha\beta}\lambda_{i}^{\alpha}\lambda_{j}^{\beta}$.
More generally an $N^{k}MHV$ amplitude has the form 
\begin{equation}
A_{n}^{N^{k}MHV}=\frac{\delta^{4}\left(p\right)\delta^{8}\left(q\right)}{\left\langle 12\right\rangle \left\langle 23\right\rangle ...\left\langle n 1\right\rangle }M_{n}^{k}
\label{Mnk}
\end{equation}
where $M_{n}^{k}$ has fermionic degree $4k$.

The amplitudes of $\mathcal{N}=4$ sYM have a remarkable symmetry
known as dual superconformal symmetry. This symmetry can be seen by
arranging the external supermomenta of an amplitude into a polygon
and writing the amplitude as a function of the vertices of this polygon.
Note that this relies on the ability to cyclically order the external
particles and is therefore only well-defined in the planar limit.
Dual superconformal symmetry then corresponds to conformal symmetry
in the dual space. In equations, the coordinates of the dual space
are defined by 

\begin{equation}
\left(x_{i}-x_{i+1}\right)^{\dot{\alpha}\alpha}=\lambda_{i}^{\alpha}\tilde{\lambda}_{i}^{\dot{\alpha}},\,\,\,\left(\theta_{i}-\theta_{i+1}\right)^{a\alpha}=\lambda_{i}^{\alpha}\eta_{i}^{a}.\label{eq:dual}\end{equation}
These coordinates automatically incorporate momentum conservation and are referred to as the coordinates of region momentum space. The dual superconformal symmetry of a scattering amplitude is the ordinary superconformal symmetry of the null polygonal Wilson loop whose vertices correspond to the points in the region momentum space of
the scattering amplitude. 

The dual superconformal symmetry of the amplitudes can be made more manifest
by writing them in terms of momentum supertwistors:
\[
\left(Z_{i}^{A},\chi_{i}^{a}\right)=\left(\lambda_{i\alpha},\mu_{i}^{\dot{\alpha}},\chi_{i}^{a}\right).\]
These transform  in the fundamental representation of the dual superconformal group $SU(2,2|4)$.

\begin{figure}[h]
\begin{center}
\includegraphics[scale=0.5]{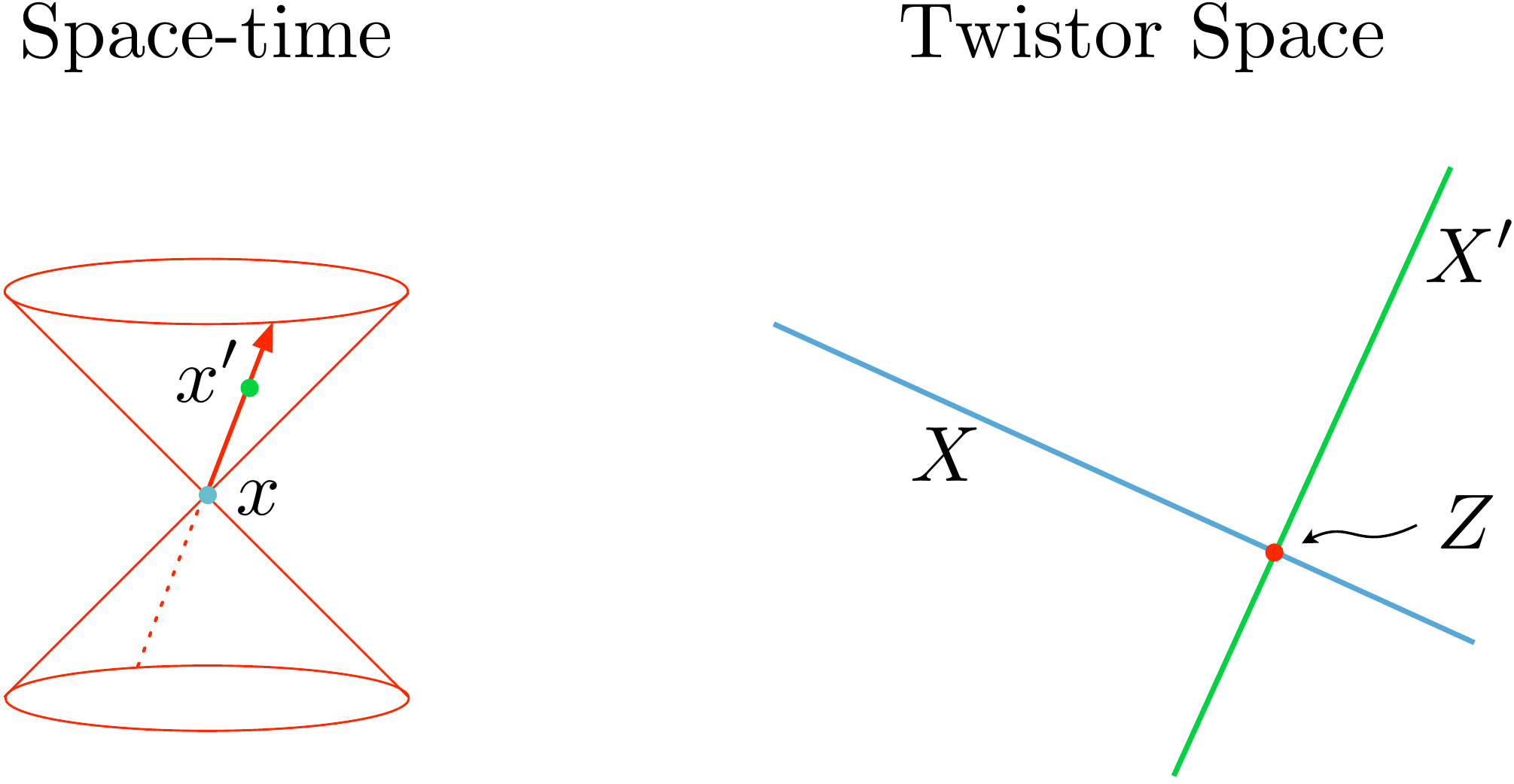}
\caption{A point in space-time corresponds to a complex projective line in twistor space.}
\label{inc}
\end{center}
\end{figure} 

The components of the momentum supertwistors can be read off from  the following 
`incidence relations' \begin{equation}
\mu_{i}^{\dot{\alpha}}=-ix_{i}^{\dot{\alpha}\alpha}\lambda_{i\alpha},\,\,\,\chi_{i}^{a}=-i\theta_{i}^{a\alpha}\lambda_{i\alpha}.\label{eq:incidence}\end{equation}
Once we specify $n$ momentum supertwistors, the corresponding
points in the dual space are given by
\[
x_{i}^{\dot{\alpha}\alpha}=i\frac{\lambda_{i}^{\alpha}\mu_{i-1}^{\dot{\alpha}}-\lambda_{i-1}^{\alpha}\mu_{i}^{\dot{\alpha}}}{\left\langle ii-1\right\rangle },\,\,\,\theta_{i}^{a\alpha}=i\frac{\lambda_{i}^{\alpha}\chi_{i-1}^{a}-\lambda_{i-1}^{\alpha}\chi_{i}^{a}}{\left\langle ii-1\right\rangle }.\]
The corresponding supermomenta can then be read off from eq \ref{eq:dual}. 

In general the incidence relations allow us to encode a point $x$ in space-time as a complex projective line $X$ in twistor space as in figure \ref{inc}.  Indeed it is clear from \eqref{eq:dual} and \eqref{eq:incidence} that $Z_i$ lies both on the line $X_i$ and $X_{i+1}$.  Otherwise said, the point  $x_i$ corresponds to the line $X_i$ passing through both $Z_i$ and $Z_{i-1}$.  
Such a projective line in
twistor space can be represented as a skew twistor
\begin{equation}
X_{i}^{AB}=\frac{Z_{i}^{[A}Z_{i-1}^{B]}}{\left\langle ii-1\right\rangle }
\label{bigX1}
\end{equation}
where we have normalized using the spinor brackets in the denominator.  Although the skew twistor is conformally invariant up to scale, its normalisation is not and requires the knowledge of the `infinity twistor' $I_{AB}$ defined by
\[
\left\langle ij\right\rangle =I_{AB}Z_{i}^{A}Z_{j}^{B},\,\,\, I_{AB}=\left(\begin{array}{cc}
\epsilon^{\alpha\beta} & 0\\
0 & 0\end{array}\right).\]
The skew twistor $I_{AB}$ is known as the infinity twistor because it corresponds to the point at infinity in conformally compactified space-time.  Given this, the distance between two points in the dual space can be written in
terms of momentum twistors as follows: 
\[
\left(x_{i}-x_{j}\right)^{2}=\frac{\left( ii-1jj-1\right) }{\left\langle ii-1\right\rangle \left\langle jj-1\right\rangle },\,\,\,\left( ijkl\right) =\epsilon_{ABCD}Z^{A}Z^{B}Z^{C}Z^{D}.\]

  \begin{figure}[h]
\begin{center}
\includegraphics[scale=0.8]{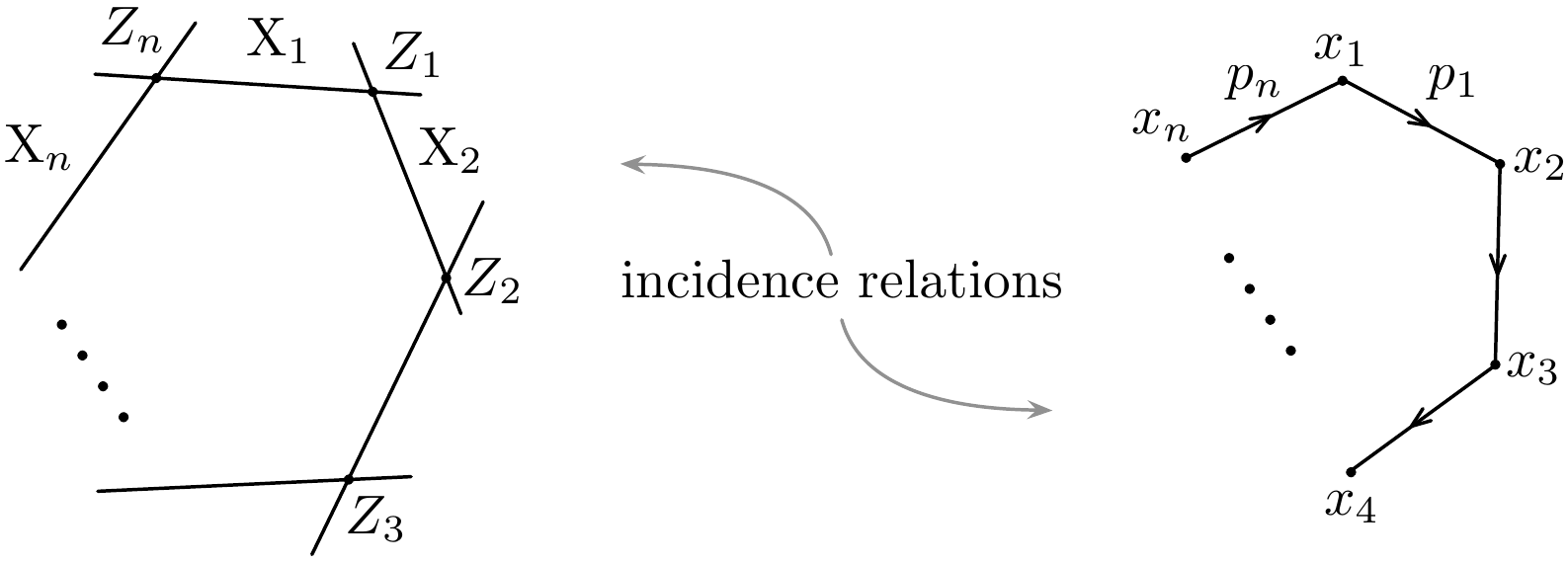}
\caption{A null polygon in space-time corresponds to a general polygon in twistor space.}
\label{poly}
\end{center}
\end{figure} 

If we arrange the momenta of a planar amplitude into a null polygon, each vertex in region momentum space
corresponds to a projective line in momentum supertwistor space, and each intersection point
in supertwistor space corresponds to a null side of the polygon in region momentum space.  Thus as depicted in figure \ref{poly} the region momentum space null polygon is dual to a generic polygon in twistor space with the vertices of one polygon corresponding to the sides of the other.

If the spacetime has Lorentzian signature, the complex conjugate of
a twistor $Z_{A}$ is a dual twistor given by $\bar{Z}_{A}=\left(\bar{\mu}^{\alpha},\bar{\lambda}_{\dot{\alpha}}\right)$.
In particular, for a momentum twistor polygon that is associated to a Lorentz real set of momenta, we must have that $\bar Z_{iA}$ is equal to $\varepsilon_{ABCD}Z_{i-1}^BZ_i^CZ_{i+1}^D$ up to scale.  
However, in Euclidean signature, complex conjugation induces the conjugation  $Z^A\rightarrow \hat{Z}_{A}\equiv\left(-\bar{\lambda}_{\dot{1}},\bar{\lambda}_{\dot{0}},-\bar{\mu}^{1},\bar{\mu}^{0}\right)$.
For more details, see for example \cite{Mason:2009qx}. Note that in Lorentzian signature, we have
\begin{equation}
X_{iAB}=\frac{\bar{Z}_{[A}\bar{Z}_{B]}}{\left[ii-1\right]}
\label{bigX2}
\end{equation}\
where $\left[ij\right]=\epsilon_{\dot{\alpha}\dot{\beta}}\bar{\lambda}_{i}^{\dot{\alpha}}\bar{\lambda}_{j}^{\dot{\beta}}$. 

\subsection{Super-Yang-Mills in twistor space}

The off-shell superfield of $\mathcal{N}=4$ sYM can be written in
terms of a $(0,1)$-form $\cA$ on supertwistor space with values in the complexified Lie algebra of the gauge group.  It  defines the $\dbar$-operator $\dbar +\cA$ on a complex vector bundle over twistor space. It has an expansion in the fermionic twistor variables as follows
\[
\mathcal{A}=g^{+}+\chi^{a}\tilde{\psi}_{a}+\frac{1}{2}\chi^{a}\chi^{b}\phi_{ab}+\epsilon_{abcd}\chi^{a}\chi^{b}\chi^{c}\left(\frac{1}{3!}\psi^{d}+\frac{1}{4!}\chi^{d}g^{-}\right)\]
where $\tilde{\psi}$ and $\psi$ are the eight  fermions,
$\phi_{ab}=-\phi_{ba}$ are the six scalars, and $g^{\pm}$ are the
positive and negative helicity states of the gluon, which are the on-shell degrees of freedom of $\mathcal{N}=4$ super-Yang-Mills. The
twistor action of $\mathcal{N}=4$ sYM is given by a holomorphic Chern-Simons
theory plus a logdet term:\[
S\left[\mathcal{A}\right]=\frac{i}{2\pi}\int D^{3|4}Z\, {\tr}\left(\mathcal{A}\wedge\bar{\partial}\mathcal{A}+\frac{2}{3}\mathcal{A}\wedge\mathcal{A}\wedge\mathcal{A}\right)+g^2 \int d^{4|8}x\log\det\left(\left.\left(\bar{\partial}+\mathcal{A}\right)\right|_{X}\right)\]
where $g^2$ is the Yang-Mills coupling and $\left.\left(\bar{\partial}+\mathcal{A}\right)\right|_{X}$
is the restriction of $\bar{\partial}+\mathcal{A}$ to the projective
line $X$ in twistor space corresponding to the point $\left(x,\theta\right)$
in spacetime. The twistor action can be interpreted as providing an expansion
about the self-dual sector of $\mathcal{N}=4$ sYM. In particular, at $\lambda=0$,
the holomorphic Chern-Simons theory describes the self-dual sector
of $\mathcal{N}=4$ sYM.  The field equations for this are simply that $(\dbar+\cA)^2=0$ so that the $\dbar$-operator is integrable and the bundle is holomorphic. By the supersymmetric Ward construction, this then gives solutions of the self-duality equations.

An on-shell field is an $\cA$ satisfying $\dbar \cA=0$ modulo $\dbar g$.  We can impose an axial gauge by choosing a reference twistor $Z_*$ and requiring that the component of $\cA$ along the lines through $Z_*$ should vanish, $\bar Z_*\cdot\cA=0$.  With this choice the cubic term in the holomorphic Chern-Simons theory vanishes, and the self-dual sector in this gauge becomes free.  
The logdet term gives the extra interactions required to extend from the self-dual sector to the full theory.   In this gauge, each term in the Taylor expansion of the
log det term corresponds to an MHV amplitude and is expressed in twistor space as an $n$-point integral over a line in twistor space that is then integrated over a real contour in space of lines, i.e., space-time,
\be\label{MHV-vertex0}
\int_{\M\times (\CP^1)^n}\hskip-.1in \frac{ \rd^{4|4}Z_A\rd^{4|4}Z_B}{\mbox{ Vol GL(2)}}\, \tr \,\prod_{i=1}^n \frac{\cA(Z(\sigma_i)) D\sigma_i}{(\sigma_{i}\,\sigma_{i+1})}\, , \quad D\sigma= (\sigma d\sigma) \, ,\quad (\sigma_i\,\sigma_j)=\sigma_{i}^0\sigma_{j}^1-\sigma_{i}^1\sigma_{j}^0\, ,
\ee
 where $Z(\sigma_i)=\sigma_i^0 Z_A+ \sigma_i^1 Z_B$ and $\sigma_{n+1}=\sigma_1$. Note that $\sigma_i=(\sigma_i^0,\sigma_i^1)$ are homogeneous coordinates for the projective line $(Z_A,Z_B)$. This yields a Feynman diagram formalism for amplitudes, the MHV diagram formalism,  whose vertices are given by off-shell extensions of the MHV amplitudes.  A more detailed description of the twistor action can be found in \cite{Adamo:2011pv}.

Such Feynman diagrams can also be used to calculate correlation functions, like the expectation value of the holomorphic Wilson loop in twistor space. As we explained in the previous subsection, if the supermomenta of a planar scattering amplitude are arranged in into a null polygon, this corresponds to a polygon in supertwistor space. The expectation value of the Wilson loop defined on this contour in twistor space is then dual to the planar scattering amplitude. Although one doesnt have a connection with which to parallelly propagate a frame for the bundle along the edges of the twistor polygon, in this holomorphic context one can nevertheless find a global holomorphic frame for the bundle on a Riemann sphere (in this perturbative context, $\cA$ is small and bundles of the Riemann sphere close to the identity are trivial).  By Liouville's theorem this frame is unique up to a global constant and so defines parallel propagation along the sides of the polygon.  Explicitly this is given by a gauge transformation $H_i$ that trivialises the $\dbar$-operator along a side $X_i$ of the twistor  polygon, satisfying 
$$
\left(\frac{\partial}{\partial\bar\sigma} +\cA(Z(\sigma))\right)H_i(\sigma)=0\, , \qquad H_i(\sigma=(1,0))=1 \qquad \mbox{ where  }\qquad Z_(\sigma)=Z_{i-1}\sigma^0+Z_i\sigma^1\, .
$$ 
This can be re-expressed as the integral equation
$$
H_i(\sigma)=I+\frac1{2\pi i}\int_{\CP^1} \frac{(\sigma_0\,\sigma)}{(\sigma_0\, \sigma')}\frac{\cA (\sigma') }{(\sigma'\, \sigma)}  H_i(\sigma')D\sigma' \quad \mbox{ where } \quad \sigma_0=(1,0)\, 
$$
which then leads to the iterative solution for $H_i$  
\be\label{sides00}
H_i(\sigma)=I+ \sum_{r=1}^\infty\frac1{(2\pi i)^r}\int _{(\CP^1)^r} \frac {(\sigma_0 \, \sigma)}{(\sigma_r \, \sigma)}\prod_{q=1}^{r}\frac {\cA(Z(\sigma_q))D \sigma_q}{(\sigma_q\,
  \sigma_{q-1})}\,  
\ee
with the product  ordered in increasing $q$ to the right. We finally define the holomorphic Wilson loop to be the trace of the holonomy of this notion of propagation around the loop 
$$
W[C]:=\tr \prod_{i=1}^n H_i(Z_i)
$$ where $H_i(Z_i)=H_i(\sigma=(0,1))$ and the product is ordered to the right in increasing $i$.   This then defines the observable for the holomorphic Wilson loop. 

In conclusion, the correlation function of the holomorphic Wilson loop in twistor space computes the loop integrand for full planar S-matrix $M_n^k$ in \eqref{Mnk}. This is more fully described in \cite{Mason:2010yk,Adamo:2011pv}.  In computing the the correlation function, we must perturbatively expand this Wilson loop in $\cA$ and take contractions of all pairs of $\cA$'s that arise at each order, replacing each pair by a propagator.  We must also include the insertion of MHV vertices from the action.  We now move on to the Feynman rules that arise from this procedure in an axial gauge following on from the choice of a reference twistor $Z_*$.

\section{Planar Wilson-loop Feynman rules in d-log form} \label{wilsonloop}
In the axial gauge there is a one to one correspondence between Feynman diagrams for the Wilson-loop correlator and MHV diagrams for the planar S-matrix, where the correspondence follows by planar duality for diagrams. In particular, vertices of the Wilson-loop Feynman diagrams correspond to faces (i.e., loops ) of the MHV diagrams.  In what follows we will use just the Feynman diagram description that arises for the Wilson-loops because this approach gives a natural geometric interpretation for the dlog form of the loop integrands. It is a simple exercise to dualize these to MHV diagrams, see \cite{Mason:2010yk} for examples. 

The Feynman rules for the holomorphic Wilson-loop lead to diagrams with vertices with all valencies from two upwards.  The loop order $L$ of the amplitude is precisely the number of vertices i.e., $L=|V|$ where $V$ is the set of vertices of the Feynamn diagram for the holomorphic Wilson loop.  The integration for the region loop momentum in the amplitude corresponds to the integral over space-time required in the MHV vertex below.  In our diagrams, we will denote an MHV vertex as a line because the support of an MHV vertex is a line in twistor space.  We will see later that the MHV degree is $k=|P|-2|V|=|P|-2L$ where $P$ is the set of propagators.   These constraints make MHV diagrams much simpler than usual Feynman diagrams.
In the planar limit we therefore write down an arbitrary planar diagram within the Wilson loop where the propagators either end on a vertex or on a side $(Z_{i-1},Z_i)$ of the Wilson-loop polygon.  The propagators  are given by
\be\label{propagator-def}
\Delta(Z,Z'):=\bar\delta^{2|4}(Z,*,Z'):=\int\frac{\rd s\rd t}{st} \bar\delta^{4|4}( Z_* + s Z+ t Z')\, .
\ee
Here $Z_*$ is the reference twistor chosen to implement an axial gauge choice, and the delta functions are defined so that for a complex variable $z=x+iy$ or fermionic variable $\eta$, 
$$
\bar\delta(z)=\delta (x) \delta(y) \rd \bar z=\frac1{2\pi i} \dbar \frac 1z \, , \qquad \delta (\eta)=\eta\, , \qquad \delta^{4|4}( Z)=\prod_{A=0}^3\bar \delta(Z^A) \prod_{a=1}^4\delta (\chi^a). 
$$
We can also define the projective delta function $\bar{\delta}^{3|4}$ by 
$$
\bar\delta^{3|4}(Z,Z')=\int_\C\frac{\rd s}{s}\bar \delta^{4|4}(Z+sZ')\, .
$$
The propagator $\Delta(Z,Z')$ is supported where $Z$, $Z_*$ and $Z'$ are collinear and has further simple poles when these points come together. This leads to the relation
$$
\dbar \Delta(Z,Z')=\bar\delta^{3|4}(Z,Z')+\bar\delta^{3|4}(Z,Z_*)+ \bar\delta^{3|4}(Z_*,Z')\, ,
$$
so that the propagator is indeed a Green's function for the $\dbar$-operator but with a couple of spurious singularities.

In a Feynman diagram for a holomorphic Wilson loop, the ends $Z$ and $Z'$ of each propagator lie either on MHV vertices, which are lines in twistor space, or on sides of the Wilson-loop in twistor space.  
An MHV vertex is located on a projective line $X$ in twistor space that corresponds to one of the (region) loop momenta $x$ that are to be integrated over.  The line will be described by a pair of points $(Z_A,Z_B)$ that lie on it up to the corresponding $GL(2,\C)$ acting on the choice $(Z_A,Z_B)$ in their linear span.   Attaching a propagator to this line involves setting $Z=Z(\sigma):=Z_A\sigma^0+Z_B\sigma^1$ for some complex parameters $(\sigma^0,\sigma^1)$  that form homogeneous coordinates on the Riemann sphere $X$.   For a vertex with $n$ propagators inserted, we multiply the $n$ propagators and must also perform the integration
\be\label{MHV-vertex}
\int_{\M\times (\CP^1)^n}\hskip-.1in \frac{ \rd^{4|4}Z_A\rd^{4|4}Z_B}{\mbox{ Vol GL(2)}}\prod_{i=1}^n \frac{D\sigma_i}{(\sigma_{i}\,\sigma_{i+1})}\, , \quad D\sigma= (\sigma d\sigma) \, ,\quad (\sigma_i\,\sigma_j)=\sigma_{i}^0\sigma_{j}^1-\sigma_{i}^1\sigma_{j}^0\, ,
\ee
where $\sigma_{n+1}=\sigma_1$, $\M$ is the chiral super Minkowski space $\R^{4|8}$ of region momentum space, and the integration must be understand as one over a real
contour inside $\C^{4|8}$ for points $x$  corresponding to the line
$(Z_A,Z_B)$.  As described above, the $GL(2)$ is that associated to
the choice of $(Z_A, Z_B)$ from within their span, and the main device
we shall use is to fix this freedom by normalizing a pair of the
$\sigma_i$, so that $Z_A$ and $Z_B$ become attachment points for propagators.   

Now consider propagators attached to a side of the polygon connecting
$Z_{i-1}$ to $Z_i$. For each propagator,  we set $Z=Z(\sigma_q)$ where
$Z(\sigma)=\sigma^0 Z_{i-1} + \sigma^1 Z_i$ in eq \ref{propagator-def}. If $r$ propagators attached to the edge, we multiply the propagators and  perform the integration 
\begin{equation}
\int _{(\CP^1)^r} \prod_{i=0}^{r}\frac {D \sigma_i}{(\sigma_i\,
  \sigma_{i+1})}\, , \quad \mbox{ where } \quad \sigma_0=(1,0)\, \quad
\mbox{ and }
\quad \sigma_{r+1}=(0,1)\,.
\label{edgefeyn} 
\end{equation}
The $\sigma=(\sigma^0,\sigma^1)$ are homogeneous coordinates on the $\CP^1$ that forms each side of the polygon.

The usual strategy as described in \cite{Mason:2010yk} would be to gauge fix the $GL(2)$ at each MHV vertex by breaking dual-conformal invariance and lining up the $\lambda$-parts of $Z_A$ and $Z_B$ with a spinor basis. This yields $\rd^{4|4}Z_A\rd^{4|4}Z_B /Vol(GL(2))=\rd^{4|8}x$ for the super region loop momentum $x$.  The remaining parameter integrals can be performed against the delta functions thus determining the parameter values in terms of invariants in the external and loop region momenta.   This then reduces to the momentum twistor MHV rules as described in \cite{Bullimore:2010pj}. See \cite{Adamo:2011dq} for a self-contained development.

Our strategy will be to fix the $GL(2)$ gauge freedom at each MHV vertex to eliminate 4 parameter integrals at each vertex.  We can then use the delta functions in the propagators to
eliminate the $\rd^{4|4}Z_A\rd^{4|4}Z_B$ integrations.  We will still
need to worry about the contours and they will be encoded in a more
indirect way in the remaining parameters. Using the relations forced by the delta functions we can of course express all the parameters in terms of the loop momenta so as to make the contours more transparent. 

\subsection{Attachment to sides, and tree-level amplitudes}
For tree-level amplitudes, there are no MHV vertices, just propagators
connected from side to side in a planar arrangement.  In this
subsection we show that the parameter integrand arising from such
insertions is always in dlog form multiplied by delta functions.

\begin{figure}
\begin{center}
\includegraphics[scale=0.17]{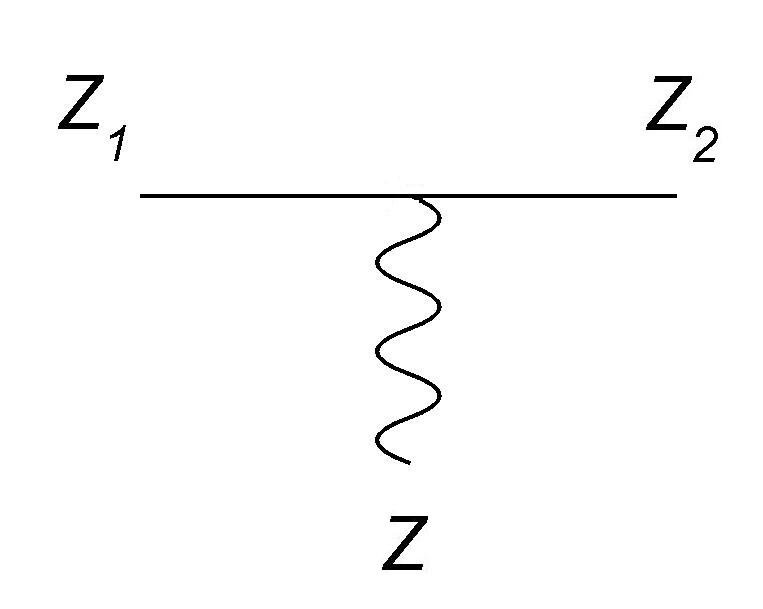}
\caption{ A twistor propagator attached
to an external edge of the Wilson loop.}
\label{fig1}
\end{center}
\end{figure}

In the simplest case each end is like
 Figure \ref{fig1}, which shows a twistor
propagator attached to an external side of the Wilson loop. Using eqs \ref{propagator-def} and \ref{edgefeyn}, this
diagram is given by  
\begin{multline}
\int_{\C^3} \frac{\rd s\rd t}{st}\frac{D\sigma
  \bar{\delta}^{4|4}\left(Z_{*}+sZ+t\sigma^0 Z_{1}+t\sigma^1 Z_{2}
  \right)}{\left(\sigma_{1}\sigma\right)\left(\sigma\sigma_{2}\right)}\\
=\int_{\C^3} \frac{\rd s}{s}\frac{\rd^2\sigma
  \bar{\delta}^{4|4}\left(Z_{*}+sZ+\sigma^{0}Z_{1}+\sigma^{1}Z_{2}
  \right)}{\sigma^0\sigma^1}\\
\label{eq:edge1}
\end{multline}
where we have redefined $t\sigma\rightarrow \sigma$ to obtain the
second formula.  If the other end is attached say to side $Z_3,Z_4$, 
with no other attached propagators, we obtain the R-invariant:
\be\label{Rinv}
[Z_*,Z_1,Z_2,Z_3,Z_4]:=\int_{\C^4} \frac{\rd^2\sigma\, \rd^2\sigma'
  \bar{\delta}^{4|4}\left(Z_{*}+\sigma^{0}Z_{1}+\sigma^{1}Z_{2}
+\sigma'^0 Z_3+\sigma'^1 Z_4
\right)}{\sigma^0\sigma^1\sigma'^0\sigma'^1}\, .
\ee
This is skew symmetric in its arguments and is a rational
function in the bosonic variables and a polynomial of degree four in
the fermionic variables
$$
[1,2,3,4,5]:=[Z_1,Z_2,Z_3,Z_4,Z_5]=\frac {\delta^{0|4}((1234)\chi_5 +
  \mbox{cyclic})}{(1234)(2345)(3451) (4512)(5123)}\, ,
$$
where $(1234)=\varepsilon_{ABCD}Z_1^AZ_2^BZ_3^CZ_4^D$ is the skew
product of the bosonic part of the four twistors $Z_1$ to $Z_4$. This can
be seen by performing the four $\sigma$-integrals against the four
bosonic delta-functions which fixes the value of say
$\sigma^0=-(*234)/(1234)$ and so on.   This is sufficient to compute 
the MHV case as  $\sum_{i<j} [*,i-1,i,j-1,j]$.  At N$^k$MHV
there are $k$ propagators and diagrams where they are attached singly to the
sides of the polygon simply correspond to products of $k$  R-invariants.

\begin{figure}
\begin{center}
\includegraphics[scale=0.17]{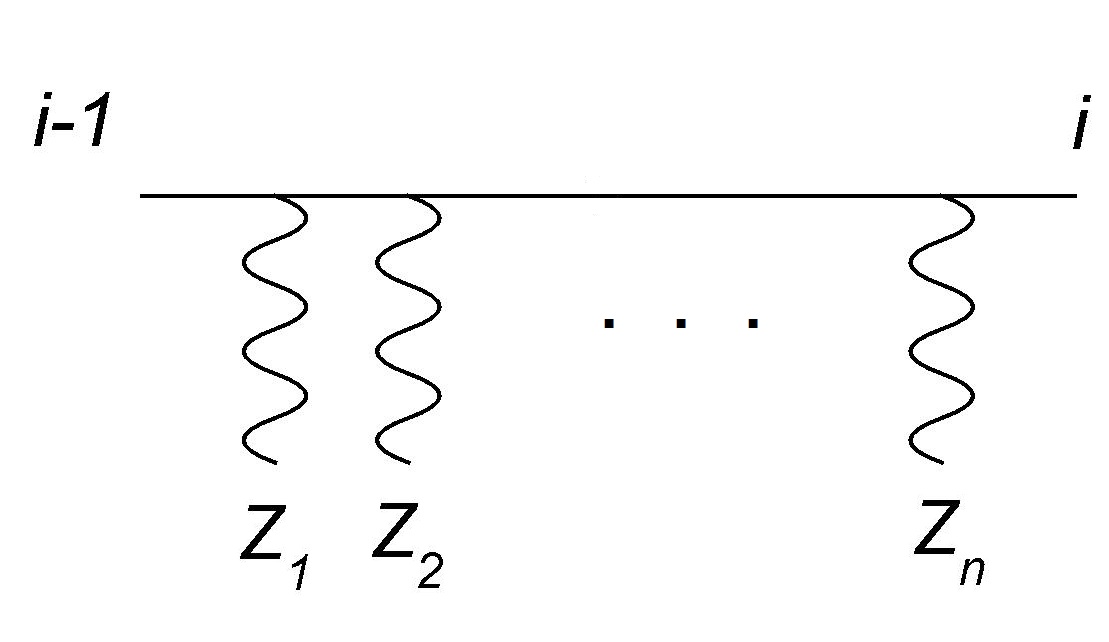}
\caption{ $n$ twistor propagators attached
to an external side of the Wilson loop.}
\label{fig2}
\end{center}
\end{figure}

Now let's consider the case of $n$ propagators attached to an external
edge, which is depicted in Figure \ref{fig2}. In this case a similar
calculation gives the diagram as
\begin{equation}
\int\frac{\rd^{2}\sigma_{1}\ldots\,\rd^{2}\sigma_{n}}
{\sigma^{1}_{1}\left(\sigma_{1}\sigma_{2}\right)\ldots \left(\sigma_{n-1}\sigma_{n}\right)\sigma^{0}_{n}}  
\prod_{r=1}^n\frac {\rd u_r}{u_r}\,\bar{\delta}^{4|4}\left(Z_{*}+\sigma_{r}^{0}Z_{i-1}+\sigma_{r}^{1}Z_{i}+u_r Z_r\right)
\,.\label{eq:edge2}\end{equation}
Letting $\sigma_{j}=t_{j}\left(1,s_{j}\right)$
for $1\leq j\leq n$, we find that $d^{2}\sigma_{j}=t_{j}ds_{j}dt_{j}$ and eq \eqref{eq:edge2} reduces to
\be\label{sides0}
\int_{\C^{3n}}
\prod_{r=1}^n  \frac{\rd s _{r}\rd t_r\rd u_r}{\left(s_{r}-s_{r-1}\right)t_r u_r}
\bar{\delta}^{4|4}\left( Z_{*}+t_r(Z_{i-1}+s_{r}Z_{i})+u_r Z_r\right)\, ,
\ee
where $s_0=0$.  This integrand is now in manifest dlog form aside from
the delta functions.   
Including the $(t_r,u_r)$ integrals in the definition of the propagators, we can simply give the reduced form for the sides as
\be\label{sides}
\int_{\C^{3n}}
\prod_{r=1}^n  \frac{\rd s _{r}}{\left(s_{r}-s_{r-1}\right)}
\bar{\delta}^{2|4}\left( Z_{*},Z_{i-1}+s_{r}Z_{i}, Z_r\right)\, .
\ee

For tree-level amplitudes these are the only ingredients, and there
are precisely as many integrals of parameters $(s_r,t_r)$ as there are
bosonic delta functions and so they are all determined in terms of
invariants of the external twistors.  All diagrams can be expressed as
products of R-invariants but perhaps with shifted entries, see
\cite{Bullimore:2010pj,Mason:2010yk} for a full discussion.

\subsection{Attachment to MHV Vertices}
\begin{figure}
\begin{center}
\includegraphics[scale=0.17]{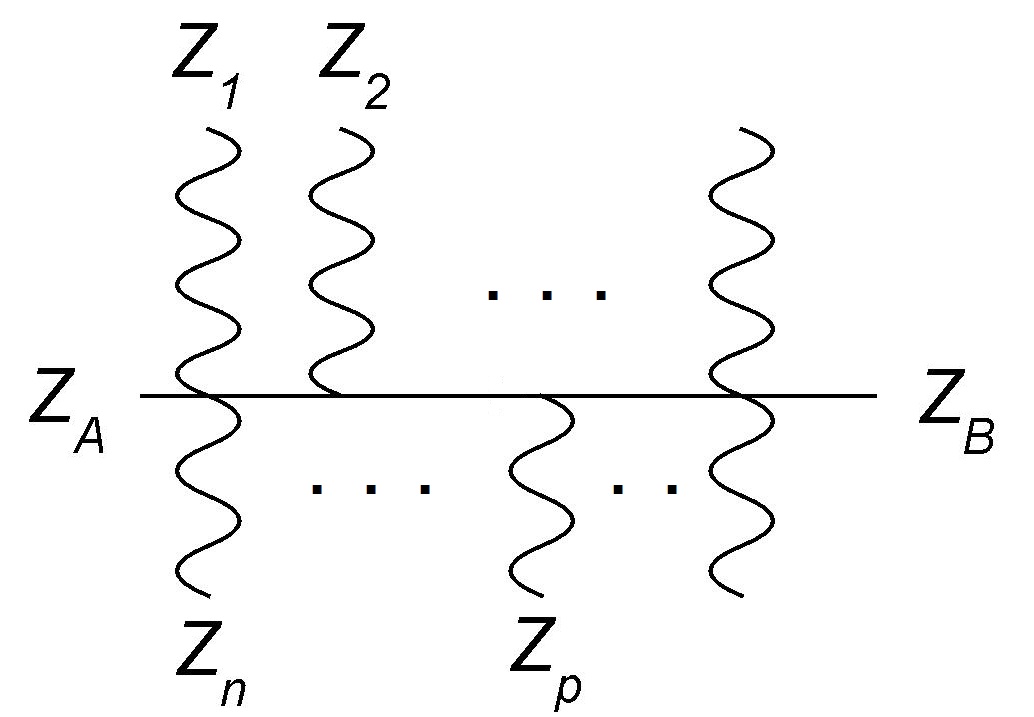}
\caption{ Twistor propagators attached to an internal vertex.}
\label{fig3}
\end{center}
\end{figure}

We will now see that all the MHV vertex integrands can be put in dlog form also.  This will be done by choosing two propagators that end on the vertex and fixing the $GL(2)$ freedom with respect to that choice.  The main distinction between the MHV vertex and an edge of the Wilson loop is that we have an additional integral over the region momentum of the loop.  Just as in the case of propagators attached to sides of the polygon, we can incorporate the parameter integral over $t$ in the definition of the propagator \eqref{propagator-def} as the scaling of the homogeneous coordinates on the Riemann sphere to obtain the following form for $n$ propagators attached to an internal vertex, as
depicted in Fig \ref{fig3} as the integral
\begin{equation}
\int \frac{\rd^{4|4}Z_{A}\rd^{4|4}Z_{B}}{GL(2)}\prod_{r=1}^n\frac{\rd^2\sigma_r \, \rd u_r}{(\sigma_r\, \sigma_{r+1})u_r}\, \bar{\delta}^{4|4}\left( Z_{*}+\sigma_{r}^{0}Z_{A}+\sigma_{r}^{1}Z_{B} +u_r Z_r\right)\, ,\label{eq:internal1}\end{equation}
where $\sigma_{n+1}=\sigma_1$.  We will use the $GL(2)$ symmetry to fix two of the $\sigma_r$ to be $(1,0)$ and $(0,1)$, say $\sigma_{1}=(1,0)$ and $\sigma_{2}=(0,1)$.  We introduce new parameter coordinates
by setting $\sigma_{j}=t_{j}\left(1,s_{j}\right)$ for $3\leq j\leq n$.  With these choices
we find that \eqref{eq:internal1} reduces to
\begin{multline}
-\int _{\C^{3n-4}}
\frac{\rd u_1\rd u_2}{u_1 u_2}
\prod_{r=3}^{n}\frac{ds_{r}dt_{r}\rd u_r}{\left(s_{r+1}-s_{r}\right)t_ru_r}  
\bar{\delta}^{4|4}\left(Z_{*}+t_{r}\left(Z_{A}+s_{r}Z_{B}\right) +u_r Z_r\right)\\ 
\rd^{4|4}Z_{A}\rd^{4|4}Z_{B} \bar{\delta}^{4|4}\left(Z_A +u_1Z_1+Z_{*}\right)\bar{\delta}^{4|4}\left(Z_B+ u_2Z_2+Z_{*}\right)
\\
=-\int _{\C^{3n-4}}
\frac{\rd u_1\rd u_2}{u_1 u_2}
\prod_{r=3}^{n}\frac{ds_{r}dt_{r}\rd u_r}{\left(s_{r+1}-s_{r}\right)t_ru_r}  
\bar{\delta}^{4|4}\left((1+t_r+s_r)Z_{*}+u_1Z_1+s_{r}u_2Z_2 +u_r Z_r\right)
\, , \label{MHV-dlog}
\end{multline}
where $s_{n+1}=0$ and to get to the last line we have integrated out $Z_A$ and $Z_B$ against the last two delta functions on the previous line to give 
\be\label{ZAZBeq}
Z_A=-Z_*-u_1 Z_1\, , \quad Z_B=-Z_*-u_2 Z_2\, .
\ee
and transformed $(t_r,u_r)\rightarrow (-t_r^{-1},u_r/t_r)$.

We can similarly gauge fix two non-adjacent propagators. For example, set $\sigma_{1}=(1,0)$, $\sigma_{p}=(0,1)$,
and let $\sigma_{j}=t_{j}\left(1,s_{j}\right)$ for $j\neq(1,p)$. Then
eq \ref{eq:internal1} then reduces to
\begin{multline}\label{MHV-dlog1}
-\int _{\C^{3n-4}} 
\prod_{r=2}^{p-1}\frac{\rd s_{r}\rd  t_r \rd u_r}{\left(s_{r}-s_{r-1}\right)t_ru_r}\prod_{r= p+1}^{n}\frac{\rd s_{r}\rd t_{r}\rd u_r}{\left(s_{r+1}-s_{r}\right)t_r u_r}\prod_{r=2,\neq p}^n\bar{\delta}^{4|4}\left(Z_{*}+t_{r}Z_{A}+t_rs_{r}Z_{B}+u_r Z_r\right)
\\
\frac{\rd u_1\rd u_p}{u_1u_p}\rd^{4|4}Z_{A}\rd^{4|4}Z_{B} \bar{\delta}^{4|4}\left(Z_{A}+Z_{*}+u_1 Z_1\right)\bar{\delta}^{4|4}\left(Z_B+Z_*+u_pZ_{p}\right)
\\=
-\int _{\C^{3n-4}} \frac{\rd u_1\rd u_p}{u_1u_p}
\prod_{r=2}^{p-1}\frac{\rd s_{r}\rd  t_r \rd u_r}{\left(s_{r}-s_{r-1}\right)t_ru_r}\prod_{r= p+1}^{n}\frac{\rd s_{r}\rd t_{r}\rd u_r}{\left(s_{r+1}-s_{r}\right)t_r u_r}\\
\prod_{r=2,\neq p}^n\bar{\delta}^{4|4}\left((1+t_r+ s_r)Z_{*}+u_1Z_{1}+s_{r}u_pZ_{p}+u_r Z_r\right)
\end{multline}
with $s_{n+1}=s_1=0$ and the second line following by integration against delta functions together with parameter redefinition as above.
Thus in both cases, the integral is in dlog form in the parameters,  together with some remaining holomorphic delta functions.  Of course we haven't really done the loop momentum integrals by integrating $\rd^{4|4}Z_A\rd^{4|4}Z_B $ against their displayed delta functions, but we have obtained new coordinates for the loop integrand. The parameters will end up being functions of the loop momenta and the external momenta once we have used all the delta functions in a diagram.

\subsection{The MHV degree and reduction to d-log form}

The fermionic integrals can all be done essentially algebraically.  There are 8 fermionic integrations required for each MHV vertex and the propagators each are homogeneous polynomials in these fermionic variables of degree 4.  The integration pulls out the coefficient of the product of the top power in the fermionic parts of $Z_A$ and $Z_B$ which has degree 8.  In order to achieve a non-zero answer, there must therefore be at least two propagators attached to each MHV vertex.   The total degree of the remaining polynomial is $4k$, where $k$ is the MHV degree.  The MHV degree is therefore easily seen to be
\be\label{MHV-deg}
k=\mbox{number of propagators}- 2\times \mbox{number of MHV vertices}=|P|-2L\, ,
\ee
where $|P|$ is the number of propagators and the loop order is, as we have already remarked, the number of vertices.  Clearly we obtain zero for $k<0$ and at MHV, $k=0$, there are no remaining fermionic parameters after integration.

In the above expressions for the MHV vertices, we have identified two propagators that are attached to each vertex in order to immediately perform the integrals  $\rd^{4|4}Z_{A}\rd^{4|4}Z_{B} \bar{\delta}^{4|4}\left(Z_A -\ldots \right)\bar{\delta}^{4|4}\left(Z_B- \ldots\right)$.  We can then use \eqref{ZAZBeq} to substitute in for $Z_A$ and $Z_B$ in terms of other twistors wherever they may appear elsewhere.  If we can do this for a diagram, then the $2L$ chosen propagators give delta functions that determine the $2L$ internal twistors $Z_A$ and $Z_B$ in terms of the external twistors and parameters.

\begin{figure}[h]
\begin{center}
\includegraphics[scale=0.17]{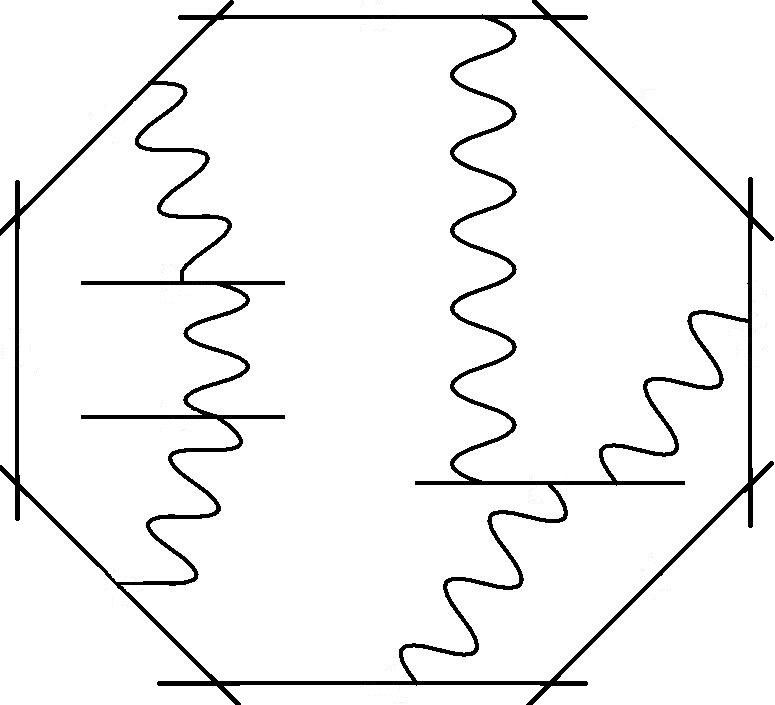}
\caption{For this vanishing diagram, one cannot associate two propagators to each vertex.}
\label{fig11}
\end{center}
\end{figure}

We claim in general that, if we are to obtain a non-zero result for the diagram, we can find an association of two propagators at each vertex to that vertex only so that we can use formulae \eqref{MHV-dlog} or \eqref{MHV-dlog1} for all the vertices in the diagram. A diagram where this is clearly not possible is
illustrated in Figure \ref{fig11} but it is easy to see that the fermionic integrals force it to vanish.
 If we look at the left part of the
diagram, we see that there are 16 fermionic integrals coming from
the two vertices but only 12 fermionic variables coming from the three
propagators. Thus it will vanish.

An inductive argument for this as follows.  It is trivially true when there is just one MHV vertex and two or more propagators.   We suppose that it is true for all diagrams with $n-1$ vertices and MHV degree $k$ that do not yield zero after fermionic integration.  A diagram  with $n$ vertices and MHV degree $k$ has one extra vertex and two extra propagators.  We can then attempt to remove a vertex, say the $n$th vertex, that is attached to the side by a propagator and that propagator. We need to identify one more propagator of those that were attached to the chosen propagator.  
We first remark that the support of the $2n+k$ propagators (determined in their $\bar\delta^{4|4}$ form)  gives $2n+k$ linear relations between the $2n$ twistors $(Z_{A_r},Z_{B_r})$, $r=1, \ldots , n$  at the MHV vertices and the $Z_i$ with coefficients in the parameters
$$
\sum_{r=1}^n A_{rI}Z_{A_r}+B_{rI}Z_{B_r}+C_{iI}Z_i=0\, , \qquad  I=1,\ldots 2n+k\, .
$$
  The fermionic integrations will yield zero if the coefficients $(A_{rI},B_{rI})$ of   $(Z_{A_r},Z_{B_r})$ as a $2n\times (2n+k)$ matrix does not have maximal rank $2n$ for generic values of the parameters. This is because there will then be some linear combination of the $Z_{A_r}$ and $Z_{B_r}$ that does not appear in the fermionic delta functions, so that the corresponding integrals of those fermionic components have no matching delta functions required to make the answer non-zero (recall that in fermionic integration $\int \eta \rd \eta=1$ but $\int \rd \eta=0$ and the fermionic $\delta$-function $\delta(\eta)=\eta$) .
Our non degeneracy assumption therefore allows us assume that $(A_{rI},B_{rI})$ does indeed have maximal rank.  Suppose that there are $r$ propagators attached to the $n$th vertex.  Since these give the only relations that involve $Z_{A_n}$ and $Z_{B_n}$, it must be the case that there exists a choice of two of them such that the rank of the remaining $(2n-2)\times (2n-2+k)$ matrix has rank $(2n-2)$ on the remaining $(Z_{A_r},Z_{B_r})$ for $r=1,\ldots ,n-1$; this is essentially a question of performing a permutation of the columns to obtain a non-degenerate row reduction.  In our scenario in which the $n$th vertex is attached to a side by a propagator, one of these can be taken to be that propagator and the other some other attached propagator.  The residual non degeneracy means that if we attach the remaining propagators at the $n$th vertex to generic points on the edges of the polygon, we will obtain a non degenerate diagram with one fewer vertex for which the assignment is now possible.  By assumption, the resulting  diagram is  one with $n-1$ vertices and will be non-zero after the remaining fermionic integrations are performed.  We can therefore by our inductive hypothesis make our assignment of pairs of propagators to vertices throughout the remainder of the diagram.

\begin{figure}[h]
\begin{center}
\includegraphics[scale=0.17]{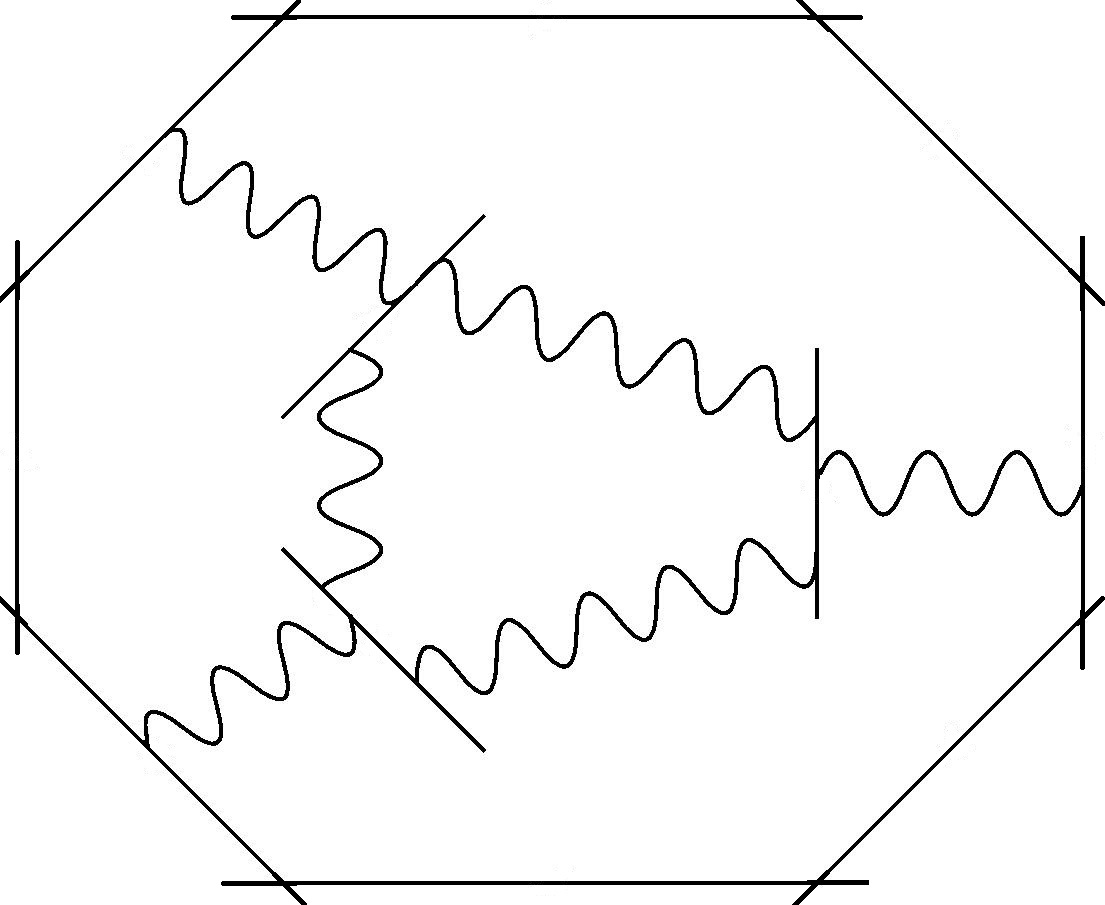}
\caption{There are two ways to assign two propagators to each vertex in this diagram.}
\label{fig12}
\end{center}
\end{figure}

For non-vanishing  diagrams with loops, there is not  generally a unique way to assign two propagators
to every vertex. In Figure \ref{fig12}, for example, there are two ways to assign
two propagators to every vertex.
Having made such a choice, for a general diagram we can simply  use the vertices in \eqref{sides0},  \eqref{MHV-dlog}, and \eqref{MHV-dlog1} and integrate out all but $k$ of the $\bar\delta^{4|4}(\ldots )$ delta functions. We also see that our procedure leaves us with 4 parameter integrals per propagator (two in the definition of the propagator in \eqref{propagator-def}, and two for the location of the end points of the propagator on the corresponding side or MHV vertex) but that we lose 4 parameter integrals against the $GL(2)$ gauge fixing at each vertex.  Thus there is a total of $4(|P|-L)=4(k+L)$ parameter integrals.  Thus $4k$ parameter integrals can in general be performed against the $4k$ remaining bosonic delta functions leaving a $4L$-form that encodes the loop integrand at arbitrary MHV degree.  

In the MHV case, $k=0$, we have already reduced to the loop integrand and it is manifestly in d-log form, consisting only of the remaining $4L$ parameter integrals as identified in \eqref{sides0}, \eqref{MHV-dlog} and \eqref{MHV-dlog1}.
We will restrict attention to this case from here on.

These parameters are generally complex but the integral must be taken over a real $4L$-dimensional slice in this complex $4L$ dimensional parameter space.  The delta functions express each pair $Z_A$ and $Z_B$ from each MHV vertex as a linear combination of otherl twistors with coefficients given by rational functions of the parameters as in \eqref{ZAZBeq}, and the corresponding  real contour will be expressed as a condition on the parameters that the line from $Z_A$ to $Z_B$ is real.  If we wish to use a Lorentz signature real slice, then these are encoded in the conditions 
\be
Z_A\cdot \bar Z_A=0=Z_B\cdot \bar Z_B= Z_A\cdot \bar Z_B\, ,
\ee
although we will also then need the Feynman $i\epsilon$ prescription.
For Euclidean signature, the complex conjugation maps a twistor to a twistor $Z\rightarrow \hat Z$ and we require that $\hat Z_A$ and $\hat Z_B$ lie in the span of $Z_A$ and $Z_B$.  We can either attempt to perform the integral directly in the given parameters, or express the parameters in terms of invariants associated with the external and loop momenta and regard the resulting differential form in dlog form as a differential $4L$ form of the $L$ region loop momenta.  In the first approach, the location of the real slice contains the external data of the amplitude, otherwise the data is embedded in the definition of the parameters in terms of the external and loop momenta.

\section{Examples} \label{examples}

Let us now apply the discussion in the previous section to the diagrams that contribute to the
1 and 2-loop MHV amplitudes as well as the 1-loop NMHV amplitude.  In what follows we will assume that we are going to be working in Lorentz signature so that to each twistor there is also associated a complex conjugate dual twistor and we will denote the contraction of a twistor and a dual twistor by $Z\cdot W$.  We will assume that the reference twistor is chosen so that $Z_*\cdot \bar Z_*=0$ and we will normalise all twistors $Z_i$  against the reference twistor so that $Z_i\cdot \bar Z_*=1$ (this in particular requires that the reference spinor is chosen to be distinct from any external momentum spinor).

\subsection{1 loop MHV} \label{1loop}

The generic 1-loop MHV diagram is depicted in Fig \ref{fig3}. 
\begin{figure}[h]
\begin{center}
\includegraphics[scale=0.14]{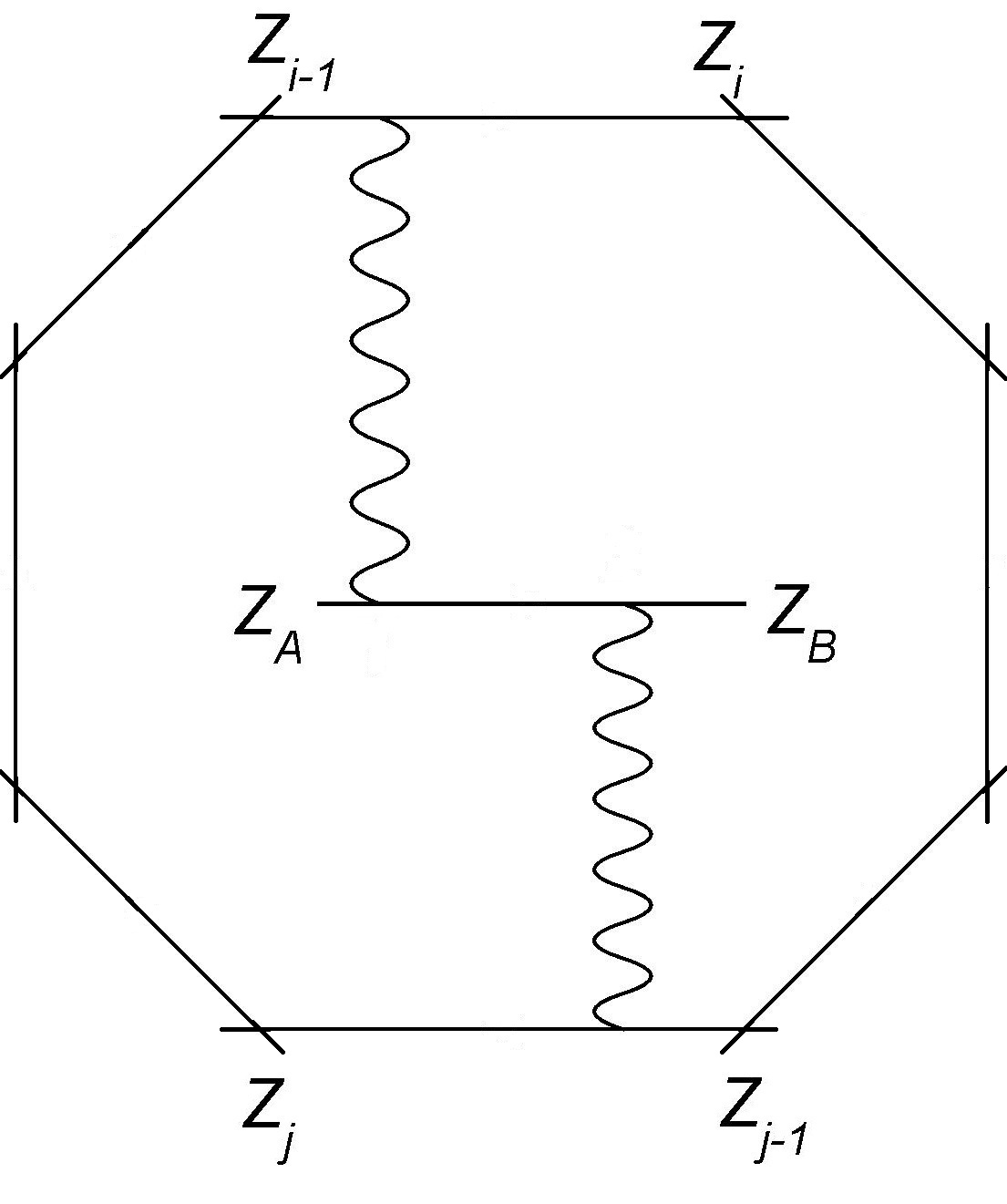}
\caption{ Twistor Wilson loop diagram for 1-loop MHV.}
\label{fig4}
\end{center}
\end{figure}
The non generic cases are when the propagators are attached to adjacent sides, which is the divergent case, or to the same side, which yields zero.  The generic case is finite and dual conformal.

Using the first line of  \eqref{MHV-dlog} at $n=2$ together with \eqref{sides0} at each insertion point on the sides, the Feynman rules give \[
\int\frac{ds_{1}}{s_{1}}\frac{dt_{1}}{t_{1}}\frac{\rd s_{2}}{s_{2}}\frac{\rd t_{2}}{t_{2}}\int \rd^{4|4}Z_{A}\rd^{4|4}Z_{B}\, \bar{\delta}^{4|4}\left(Z_{A}-Z_{*}-s_{1}Z_{i-1}-t_{1}Z_{i}\right)\bar{\delta}^{4|4}\left(Z_{B}-Z_{*}-s_{2}Z_{j-1}-t_{2}Z_{j}\right).\]
Doing $Z_{A}$ and $Z_{B}$ integrals against delta functions then
leaves\[
\int\frac{ds_{1}}{s_{1}}\frac{dt_{1}}{t_{1}}\frac{ds_{2}}{s_{2}}\frac{dt_{2}}{t_{2}}\]
with
\be\label{kermit-parameters1}
Z_A=Z_*+s_1 Z_{i-1}+t_1 Z_i\, , \qquad Z_B=Z_*+s_2Z_{j-1} + t_2 Z_j\, .
\ee
In fact, in what follows it will be convenient to perform a rational transformation of the parameter variables that preserves the form of the integrand. Letting $\left(s_{1},t_{1}\right)=-\frac{i}{s_{0}(1+s)}\left(1,s\right)$ and $\left(s_{2},t_{2}\right)=-\frac{i}{t_{0}(1+t)}\left(1,t\right)$, we have
\be\label{kermit-parameters}
Z_{A}=is_{0}Z_{*}+\frac{1}{1+s}\left(Z_{i-1}+sZ_{i}\right)\, , \qquad Z_{B}=it_{0}Z_{*}+\frac{1}{1+t}\left(Z_{j-1}+tZ_{j}\right)\, .
\ee
This has the benefit that the integral is
$$
\int \rd\log s_0\rd \log s\rd \log t_0\rd\log t
$$ 
but now $Z_A$ and $Z_B$ are normalised against $\bar Z_*$ and formulae for the contour will be  simpler.

\begin{figure}[h]
\begin{center}
\includegraphics[scale=0.15]{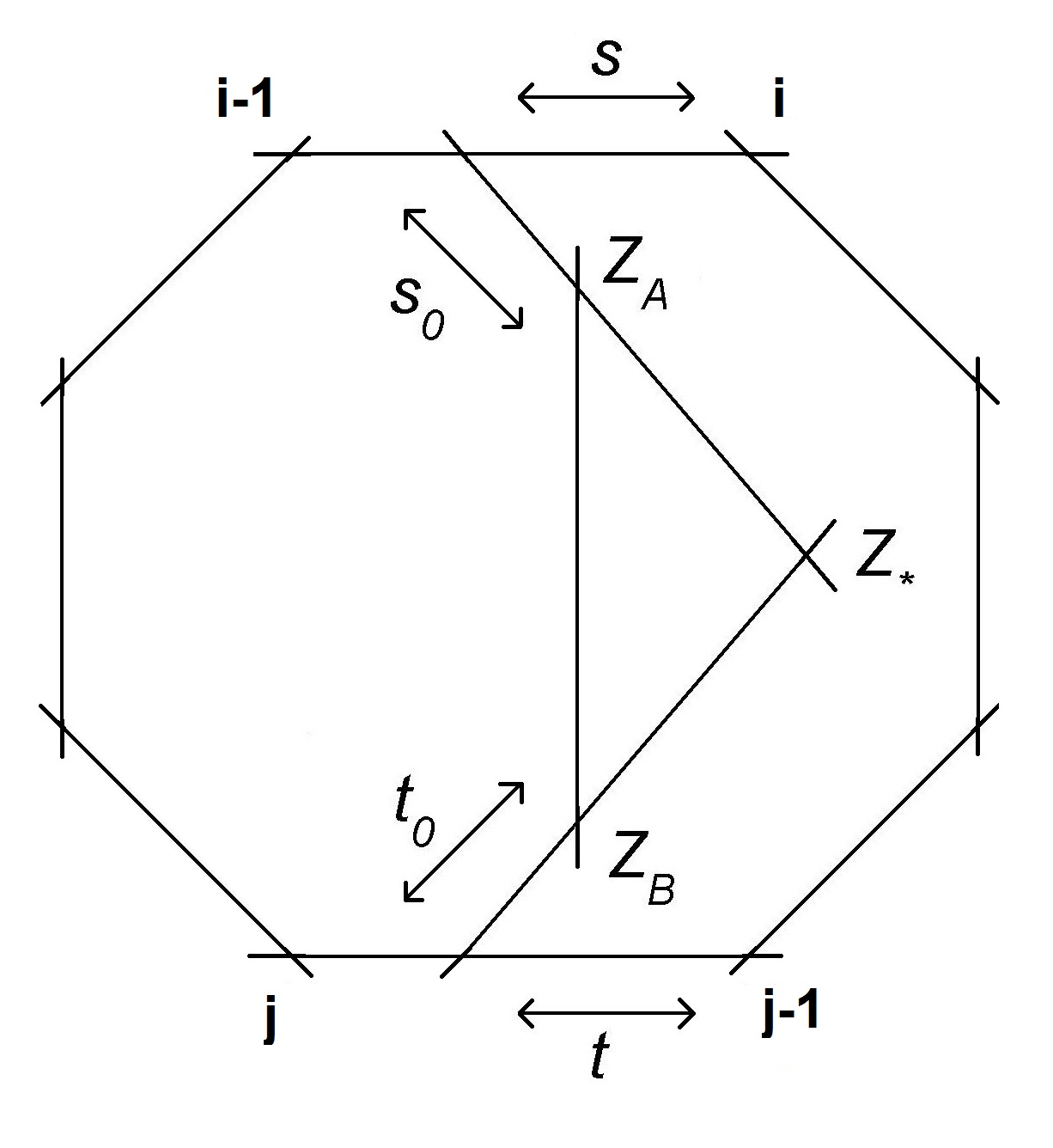} 
\caption{Geometric interpretation of parameters for Kermit integral. }
\label{s0t0}
\end{center}
\end{figure}

The parameters can be understood geometrically in twistor space as follows.  Given the line $X_0$ on which the MHV vertex is supported, let  $Z_A$ be its intersection with the plane through $Z_{i-1}$, $Z_i$ and $Z_*$, and $Z_B$ its intersection with the plane through $Z_{j-1}$, $Z_j$ and $Z_*$.  Then $s$ parametrizes the intersection point $Z_{i-1}+sZ_i$ of the line through $Z_*$ and $Z_A$ with $X_i$, and $s_0$ parametrizes the position of $Z_A$ along the line from $Z_*$ to $Z_{i-1}+sZ_i$ and similarly for $Z_B$; see Figure \ref{s0t0}.  

One approach to evaluating the above integral is to re-express it in ordinary (region) momentum space coordinates.  This can be done by writing $Z_*=(0,\xi^{\dot\alpha})$, $Z_i=(\lambda_{i\alpha}, -ix_i^{\dot \alpha\alpha}\lambda_{i\alpha})$ and $Z_A=(\lambda_{A\alpha}, -ix_0^{\dot \alpha\alpha}\lambda_{A\alpha})$, where $x_0$ is the loop region momentum.  Then \eqref{kermit-parameters}   gives
$$
Z_A:=(\lambda_{A}, -ix_0|\lambda_{A}\ra)= s_0(0,\xi)+ \frac1{1+s}(\lambda_{i-1}+s\lambda_{i},-ix_i|\lambda_{i-1\, }+s\lambda_{i}\ra)
$$ 
and after some algebra using also the normalization conditions $\la \lambda_i\, \bar \xi\ra=1=\la\lambda_A\,\bar\xi\ra$  we find
\be\label{}
s_0=-\frac{x_{0i}^2}{[\xi|x_{0i}|\bar\xi\ra}\, , \quad t_0=-\frac{x_{0j}^2}{[\xi|x_{0j}|\bar\xi\ra}\, , \quad s= -\frac{[\xi|x_{0i}|\lambda_{i-1}\ra }{[\xi|x_{0i}|\lambda_{i}\ra}\, ,\quad t= -\frac{[\xi|x_{0j}|\lambda_{j-1}\ra }{[\xi|x_{0j}|\lambda_{j}\ra}\, ,
\ee
where $x_{0i}=x_0-x_i$ etc..  These are in fact dual conformally invariant expressions but for the choice of the reference twistor
\begin{eqnarray}\label{svsx}
s_0&=&\frac{i}{2}\frac{X_0\cdot X_i}{Z_{*}\cdot X_0\cdot X_{i}\cdot \bar Z_{*}}\, , \quad t_0=\frac{i}{2}\frac{X_0\cdot X_j}{Z_*\cdot X_0\cdot X_j\cdot \bar Z_*}\, , \nonumber \\
 s&=& -\frac{Z_*\cdot X_0\cdot Z_{i-1} }{Z_*\cdot X_0\cdot Z_i}\, ,\quad t= -\frac{Z_*\cdot X_0 \cdot Z_{j-1} }{Z_*\cdot X_0\cdot Z_{j}}\, ,
\end{eqnarray}
where we have introduced the notation $Z_i\cdot X_0 \cdot X_j \cdot \bar Z_*:=Z^A_iX_{0AB}X_j^{BC}\bar Z_{*C}$ and we have used eqs \ref{bigX1} and \ref{bigX2}. 
With this, the integrations can be performed in the $x_0$ region momentum variables but with the benefit of having an integrand that is locally exact so that Stokes theorem can be repeatedly applied on the complement of the singularities of the integrand in order to perform the integration.  This can be done in either Euclidean or Lorentzian signature.  However, in the latter case we note that we will require an $i\epsilon$ prescription, $x_{0i}^2\rightarrow x_{0i}^2+i\epsilon$. 


\subsubsection{Contour} \label{contour}
Instead of performing the integral in region momentum coordinates, we can instead attempt to do so in the parameter coordinates.  However, these complex parameters (or at least $s$ and $t$ are, as we shall see) and the integrand is sufficiently singular that it will not be independent of the choice of contour.  The contour is given by the  condition that $x_0$ is real.  Although we could  use Euclidean signature,  we will here only explore the case in which $x_0$ is Lorentz real.  This is encoded
in the condition that the line from $Z_A$ to $Z_B$ lies in $Z\cdot \bar Z=0$ which is equivalent to
$$
0= Z_A\cdot \bar Z_A, ,  \qquad 0=Z_B\cdot \bar Z_B\, , \qquad 0=Z_A\cdot \bar Z_B\, .
$$
The reality conditions then  give
\be\label{real0}
0= Z_A\cdot \bar Z_A= i(s_0-\bar s_0)\, ,  \qquad 0=Z_B\cdot \bar Z_B=i(t_0-\bar t_0)\, , 
\ee
and 
\be\label{real1}
 0=Z_A\cdot \bar Z_B= iv (1+s)(1+\bar t) +  
  a_{i-1\, j-1} + s  a_{i\, j-1} +\bar t a_{i-1\, j} + s\bar t a_{ij}\, ,
\ee
where we have set $a_{ij}=Z_i\cdot\bar Z_j$ etc.\ and $v= s_0-\bar t_0$. In obtaining these equations, we noted that $\bar{Z}_*\cdot Z_*=\bar{Z}_{i}\cdot Z_i=\bar{Z}_{i-1}\cdot Z_i=0$ and $\bar{Z}_{i-1}\cdot Z_*=1$ and similarly for $i\leftrightarrow j$. 

Then \eqref{real1} gives the contour for the $(s,t)$ integral as 
 \begin{equation}
s=-\frac{\bar{t}\left(a_{i-1 j}+iv\right)+a_{i-1 j-1}+iv}{\bar{t}\left(a_{ij}+iv\right)+a_{i j-1}+iv}\label{stconstraint}\end{equation}
where $v=s_{0}-t_{0}$.  Thus the 1-loop MHV amplitude reduces to the  integral
\begin{equation}
\int\frac{ds_{0}}{s_{0}}\frac{dt_{0}}{t_{0}}\frac{ds}{s}\frac{dt}{t}
\label{kermitfinal}
\end{equation}
over the contour given in \eqref{real0} and \eqref{stconstraint}. The $(s_0,t_0)$ integrals have real poles which require regularisation, and this is done via the Feynman $i\epsilon$ prescription as described above. 

The result of the Kermit integral should depend only on the four quantiies $(a_{ij},a_{i-1\, j},a_{i\, j-1},a_{i-1\, j-1})$.  
In appendix \ref{reverse} we show that the following expression gives the correct symbol of the 1-loop MHV amplitude: 
\be\label{kermit-fin}
\Li_{2}\left(\frac{ia_{ij-1}}{v_{*}}\right)+\Li_{2}\left(\frac{ia_{i-1j}}{v_{*}}\right)-\Li_{2}\left(\frac{ia_{ij}}{v_{*}}\right)-\Li_{2}\left(\frac{ia_{i-1j-1}}{v_{*}}\right) + \mbox{c.c.}\, 
\ee
where
\[
v_{*}=\frac{X_{i}\cdot X_{j}}{i\bar{Z}_{*}\cdot X_{i}\cdot X_{j}\cdot Z_{*}}=i\left(\frac{a_{ij-1}a_{i-1j}-a_{ij}a_{i-1j-1}}{a_{i-1j}-a_{ij}-a_{i-1j-1}+a_{ij-1}}\right).
\]
In a future publication \cite{upcoming}, we will obtain \eqref{kermit-fin} by performing the integral in eq \eqref{kermitfinal} using an $i\epsilon$ prescription.  

\subsection{2 loop MHV}
\begin{figure}[h]
\begin{center}
\includegraphics[scale=0.25]{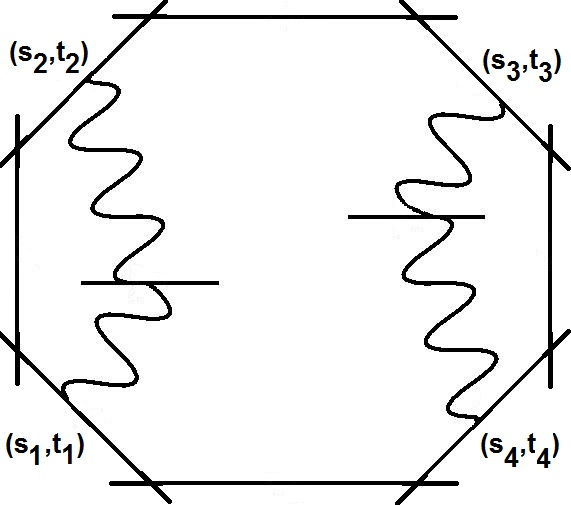} 
\caption{First topology type for the 2-loop MHV amplitude. }
\label{fig6}
\end{center}
\end{figure}

At two loops we now have two MHV vertices, and we must also have four propagators at MHV.   
There are only two topologies for the 2-loop MHV diagrams that give rise to non-zero contributions.  The first is figure \ref{fig6}

This is essentially a product of two one-loop MHV amplitudes and the discussion of the previous subsection applies directly to give an integrand in the form of eight dlogs
\begin{equation}
\int\frac{ds_{1}}{s_{1}}\frac{dt_{1}}{t_{1}}\frac{ds_{2}}{s_{2}}\frac{dt_{2}}{t_{2}}\frac{ds_{3}}{s_{3}}\frac{dt_{3}}{t_{3}}\frac{ds_{4}}{s_{4}}\frac{dt_{4}}{t_{4}}\label{eq:alpha}\end{equation}
where $(s_1,t_1,s_2,t_2)$ correspond to the parameters associated with the left hand part of the diagram and $(s_3,t_3,s_4,t_4)$ the right hand part.  The contour will be a cartesian product of that for the one-loop case (and indeed the integrand factorizes appropriately also) so the final answer is the product of the two one loop answers.   
 
\begin{figure}[h!]
\begin{center}
\includegraphics[scale=0.25]{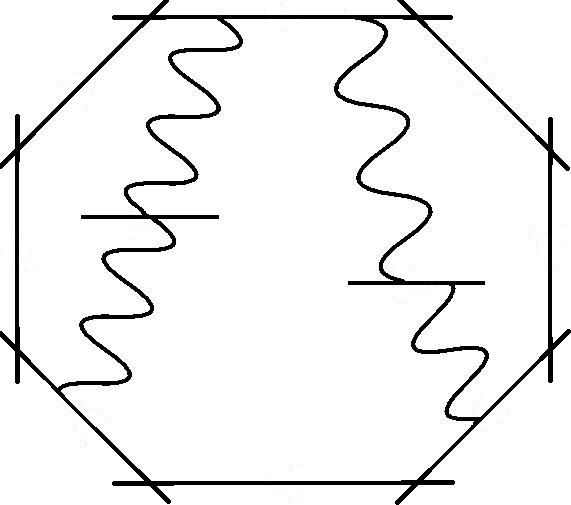}\qquad \includegraphics[scale=0.25]{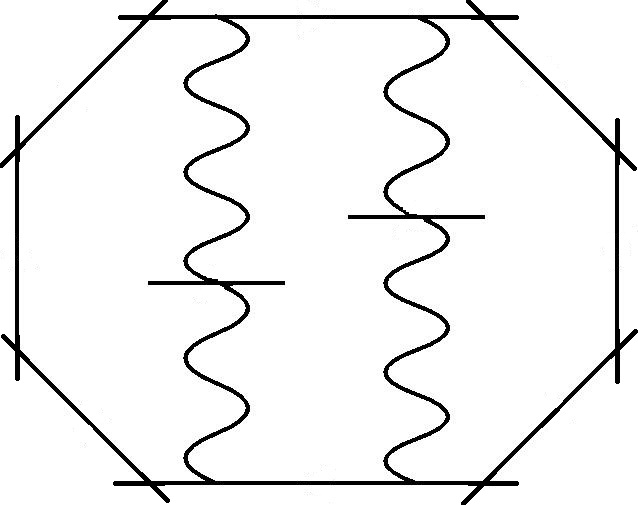}
\caption{Boundary versions of the diagram in fig \ref{fig6}.}
\label{fig7}
\end{center}
\end{figure}

 However, this topology type does have boundary versions when one or two pairs  of  propagators  end up on the same side as in figure \ref{fig7}. 
  In this case we must use \eqref{sides0} or \eqref{sides} for the treatment of the sides with two propagators which leads to a shift in the denominator of one of the parameters associated to that side, so that for the left hand diagram we obtain
  \be
  \int\frac{ds_{1}}{s_{1}}\frac{dt_{1}}{t_{1}}\frac{ds_{2}}{s_{2}}\frac{dt_{2}}{t_{2}}\frac{ds_{3}}{s_{3}-s_2}\frac{dt_{3}}{t_{3}}\frac{ds_{4}}{s_{4}}\frac{dt_{4}}{t_{4}}\label{eq:beta}\end{equation}  
  for the left hand diagram or
   \be
  \int\frac{ds_{1}}{s_{1}-s_4}\frac{dt_{1}}{t_{1}}\frac{ds_{2}}{s_{2}}\frac{dt_{2}}{t_{2}}\frac{ds_{3}}{s_{3}-s_2}\frac{dt_{3}}{t_{3}}\frac{ds_{4}}{s_{4}}\frac{dt_{4}}{t_{4}}\label{eq:beta}\end{equation}  
  for the right hand one.  The contours are given as before but the integrands no longer factorize.  (The integrands can be put back into unshifted dlog form, by redefining $s_3-s_2$ as $s_3$, but this will then mean that the definition of the contour will no longer factorize.) 

\begin{figure}[h]
\begin{center}
\includegraphics[scale=0.16]{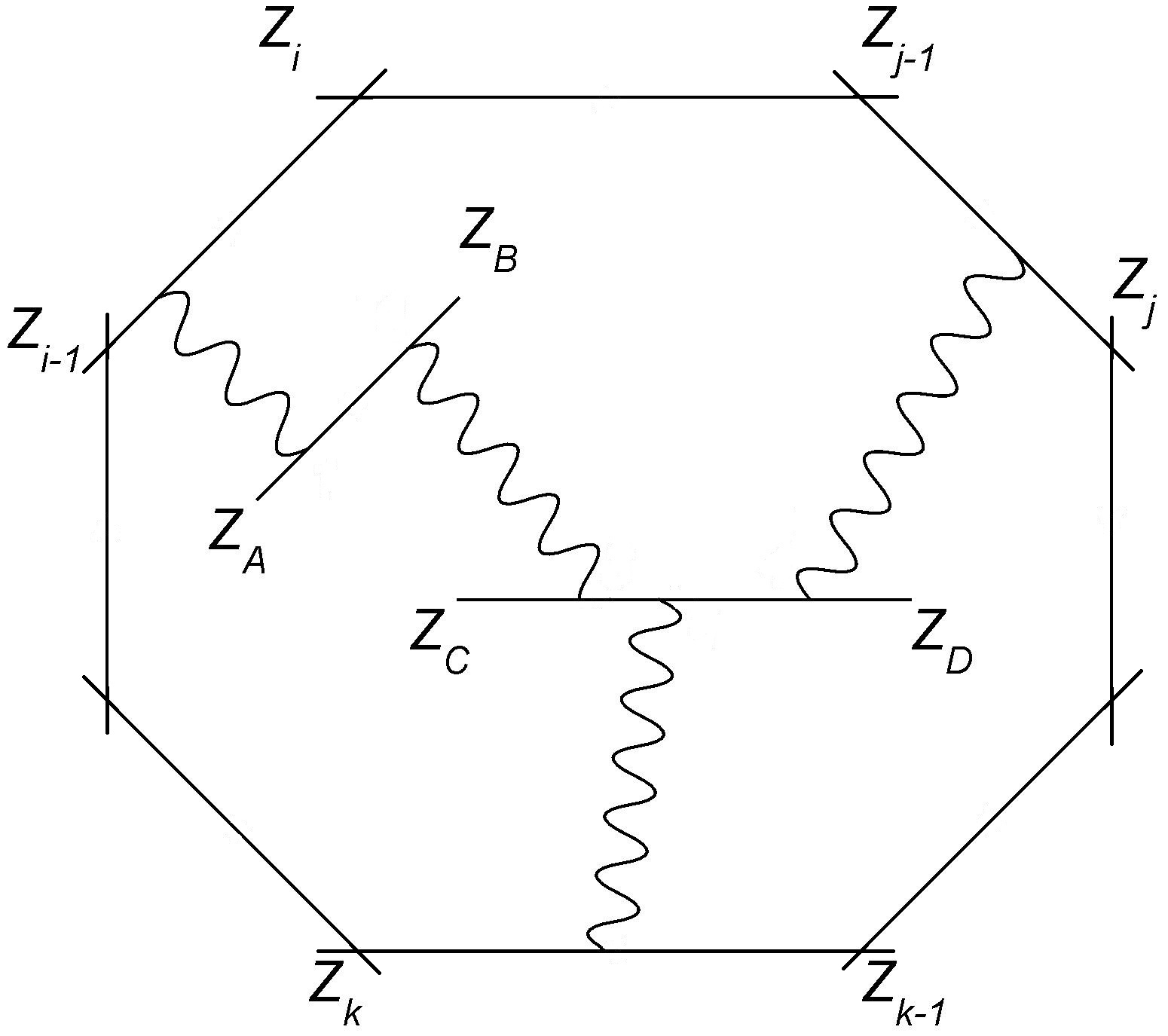}
\caption{Second topology type for the 2-loop MHV amplitude.}
\label{fig9}
\end{center}
\end{figure}

The second topological type that gives a nontrivial answer is given in generic form in figure \ref{fig9}.
There is just one way of assigning the pairs of propagators to each vertex. Using the first line of  \eqref{MHV-dlog} for $n=2$ and $n=3$ as well as \eqref{sides0} at each insertion point on the sides, the Feynman rules give 
\begin{multline}
\int\frac{\rd s_{1}}{s_{1}}\frac{\rd t_{1}}{t_{1}}\frac{\rd s_{2}}{s_{2}}\frac{\rd t_{2}}{t_{2}}\frac{\rd s_{3}}{s_{3}}\frac{\rd t_{3}}{t_{3}} \frac{\rd s_4}{s_4}\frac{\rd t_4}{t_4} \rd^{4|4}Z_{A}\rd^{4|4}Z_{B}\rd^{4|4}Z_{C}\rd^{4|4}Z_{D}
\\
\bar{\delta}^{4|4}\left(Z_A- Z_{*}-s_{1}Z_{i-1}-t_{1}Z_{i}\right)\bar{\delta}^{4|4}\left(Z_B-Z_{*}-s_2 Z_{C}-t_2 Z_{D}\right)\\
\bar{\delta}^{4|4}\left(Z_C-Z_{*}-s_{3}Z_{j-1}-t_{3}Z_{j}\right)\bar{\delta}^{4|4}\left(Z_D-Z_*-s_{4}Z_{k-1}-t_{4}Z_{k}\right).
\label{eq:mhv2}
\end{multline}
After integrating out the delta functions we are left  as before with
\begin{equation}
\int\frac{\rd s_{1}}{s_{1}}\frac{\rd t_{1}}{t_{1}}\frac{\rd s_{2}}{s_{2}}\frac{\rd t_{2}}{t_{2}}\frac{\rd s_{3}}{s_{3}}\frac{\rd t_{3}}{t_{3}}\frac{\rd s_{4}}{s_{4}}\frac{\rd t_{4}}{t_{4}}
\label{2loop}
\end{equation}
but now the contour is determined by the reality of the lines $(Z_A,Z_B)$ and $(Z_C,Z_D)$ when given by the equations
\[
Z_A= Z_{*}+s_{1}Z_{i-1}+t_{1}Z_{i}\, , \quad Z_B=Z_{*}+s_2 Z_{C}+t_2 Z_{D}\, , 
\]
\begin{equation}
Z_C=Z_{*}+s_{3}Z_{j-1}+t_{3}Z_{j}\, ,\quad Z_D=Z_*+s_{4}Z_{k-1}+t_{4}Z_{k}\, .
\label{4twist}
\end{equation}

\begin{figure}[h]
\begin{center}
\includegraphics[scale=0.17]{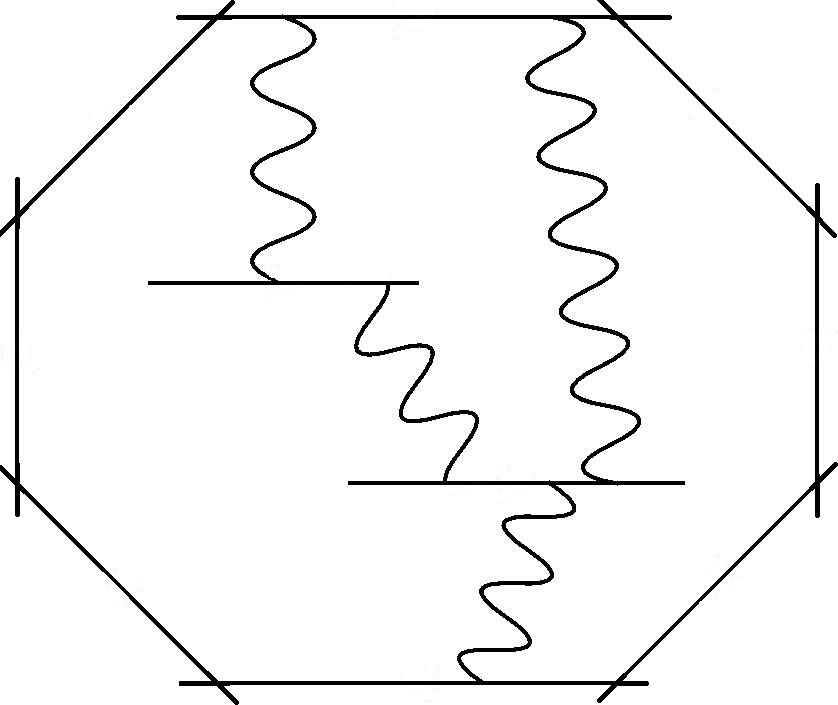}
\caption{Boundary version of the diagram in fig \ref{fig9}.}
\label{fig10}
\end{center}
\end{figure}

This diagram also has a corresponding boundary version  with two propagators ending on the same side, as depicted in figure \ref{fig10}. 
As before, this  leads to a shifted version of the above integrand
\begin{equation}
\int\frac{\rd s_{1}}{s_{1}}\frac{\rd t_{1}}{t_{1}}\frac{\rd s_{2}}{s_{2}}\frac{\rd t_{2}}{t_{2}}\frac{\rd s_{3}}{s_{3}-s_1}\frac{\rd t_{3}}{t_{3}}\frac{\rd s_{4}}{s_{4}}\frac{\rd t_{4}}{t_{4}}
\label{2loop-shift}
\end{equation}
but with the same contour (except that now $i=j$).

\begin{figure}[h]
\begin{center}
\includegraphics[scale=0.16]{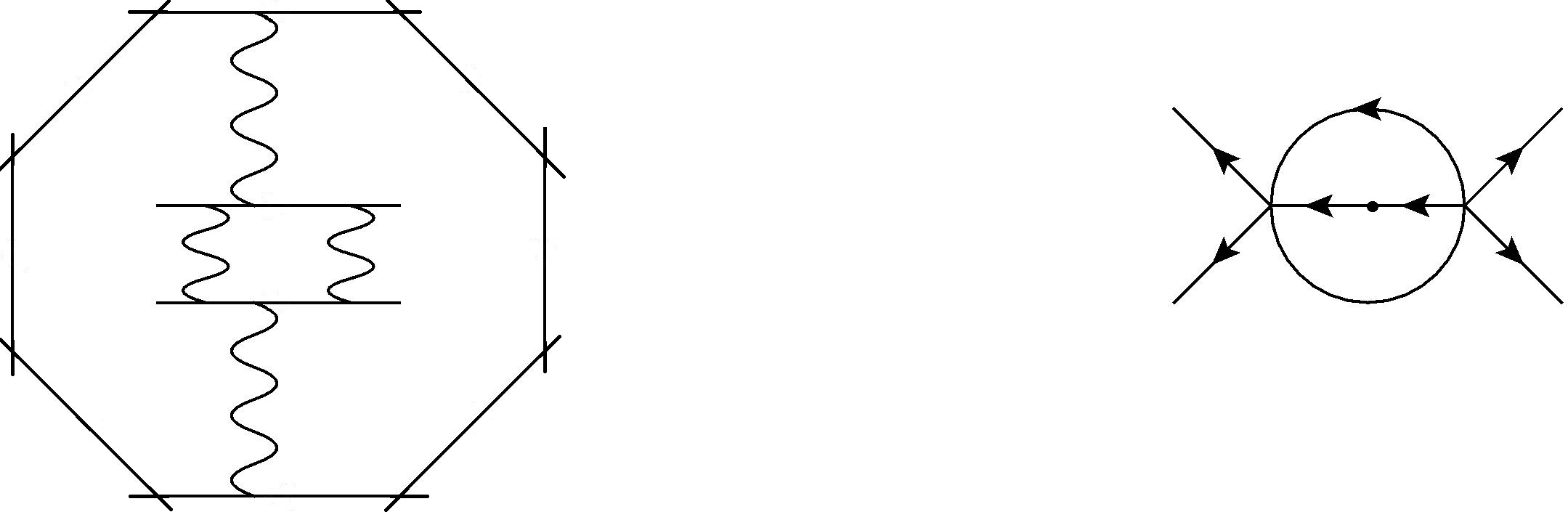}
\caption{ Diagram for vanishing 2-loop MHV contribution and its dual MHV diagram.}
\label{fig5}
\end{center}
\end{figure}

A third topological type is given by the diagram in Figure \ref{fig5} 
but this can immediately be seen to vanish since its dual MHV diagram contains an on-shell 2pt MHV vertex. 

\subsubsection{Contour}

We will now derive the contour for the two-loop diagram in Figure \ref{fig9}. As explained in the previous subsection, this is essentially the only nontrivial diagram which appears at two loops. Up to shifts in the integration variables, the other diagrams reduce to figure \ref{fig9} or to the square of the 1-loop MHV diagram. The reality constraints for this diagram are given by

\[
Z_{A}\cdot\bar{Z}_{A}=Z_{B}\cdot\bar{Z}_{B}=Z_{C}\cdot\bar{Z}_{C}=Z_{D}\cdot\bar{Z}_{D}=Z_{A}\cdot\bar{Z}_{B}=Z_{C}\cdot\bar{Z}_{D}=0.\]
Performing a rational transformation of the variables in eq \ref{2loop}, the momentum
twistors in eq \ref{4twist} can be parameterized as follows:

\[
Z_{A}=is_{0}Z_{*}+\frac{1}{s+1}\left(Z_{i-1}+sZ_{i}\right),\,\,\, Z_{B}=it_{0}Z_{*}+\frac{1}{1+t}\left(Z_{C}+tZ_{D}\right)\]
\[
Z_{C}=iu_{0}Z_{*}+\frac{1}{1+u}\left(Z_{j-1}+uZ_{j}\right),\,\,\, Z_{D}=iv_{0}Z_{*}+\frac{1}{1+v}\left(Z_{k-1}+vZ_{k}\right)\]
Note that the integrand still has the dlog form in terms of the new
integration variables:\begin{equation}
\int{\normalcolor \rd\log s_{0}\rd\log s\rd\log t_{0}\rd\log t\rd\log u_{0}\rd\log u\rd\log v_{0}\rd\log v}.\label{eq:2loopdlog}\end{equation}
The constraints $Z_{A}\cdot\bar{Z}_{A}=Z_{C}\cdot\bar{Z}_{C}=Z_{D}\cdot\bar{Z}_{D}=Z_{C}\cdot\bar{Z}_{D}=0$
are very similar to the ones we encountered when deriving the contour
for the 1-loop MHV amplitude. In particular, these constraints imply
that \begin{equation}
s_{0}=\bar{s}_{0},\,\,\, u_{0}=\bar{u}_{0},\,\,\, v_{0}=\bar{v}_{0}\label{eq:2loop1}\end{equation}
\begin{equation}
u=-\frac{\left(a_{j-1k}+iw\right)\bar{v}+a_{j-1k-1}+iw}{\left(a_{jk}+iw\right)\bar{v}+a_{jk-1}+iw},\,\,\, w=u_{0}-v_{0}.\label{eq:2loop2}\end{equation}
Furthermore, given that $Z_{C}\cdot\bar{Z}_{C}=Z_{D}\cdot\bar{Z}_{D}=Z_{C}\cdot\bar{Z}_{D}=0$,
the constraint $Z_{B}\cdot\bar{Z}_{B}=0$ implies that \begin{equation}
t_{0}=\bar{t}_{0}.\label{eq:2loop3}\end{equation}
Finally, the constraint $Z_{A}\cdot\bar{Z}_{B}$ implies that 
\begin{multline}
s_{0}-t_{0}-u_{0}+\left(s_{0}-t_{0}-v_{0}\right)\bar{t}
=
\frac{i}{1+s}\left[\frac{1}{1+\bar{u}}\left(a_{i-1j-1}+\bar{u}a_{i-1j}+s\left(a_{ij-1}+\bar{u}a_{ij}\right)\right) \right. \\
\left. +
\frac{\bar{t}}{1+\bar{v}}\left(a_{i-1k-1}+\bar{v}a_{i-1k}+s\left(a_{ik-1}+\bar{v}a_{ik}\right)\right)\right].\label{eq:2loop4}
\end{multline}

In summary, the 2-loop amplitude in Figure \ref{fig9} is given by the integral
in eq \ref{eq:2loopdlog} and the contour for this integral is defined by eqs \ref{eq:2loop1}-\ref{eq:2loop4}.

\subsection{NMHV at 1-loop}
\begin{figure}[h]
\begin{center}
\includegraphics[scale=0.16]{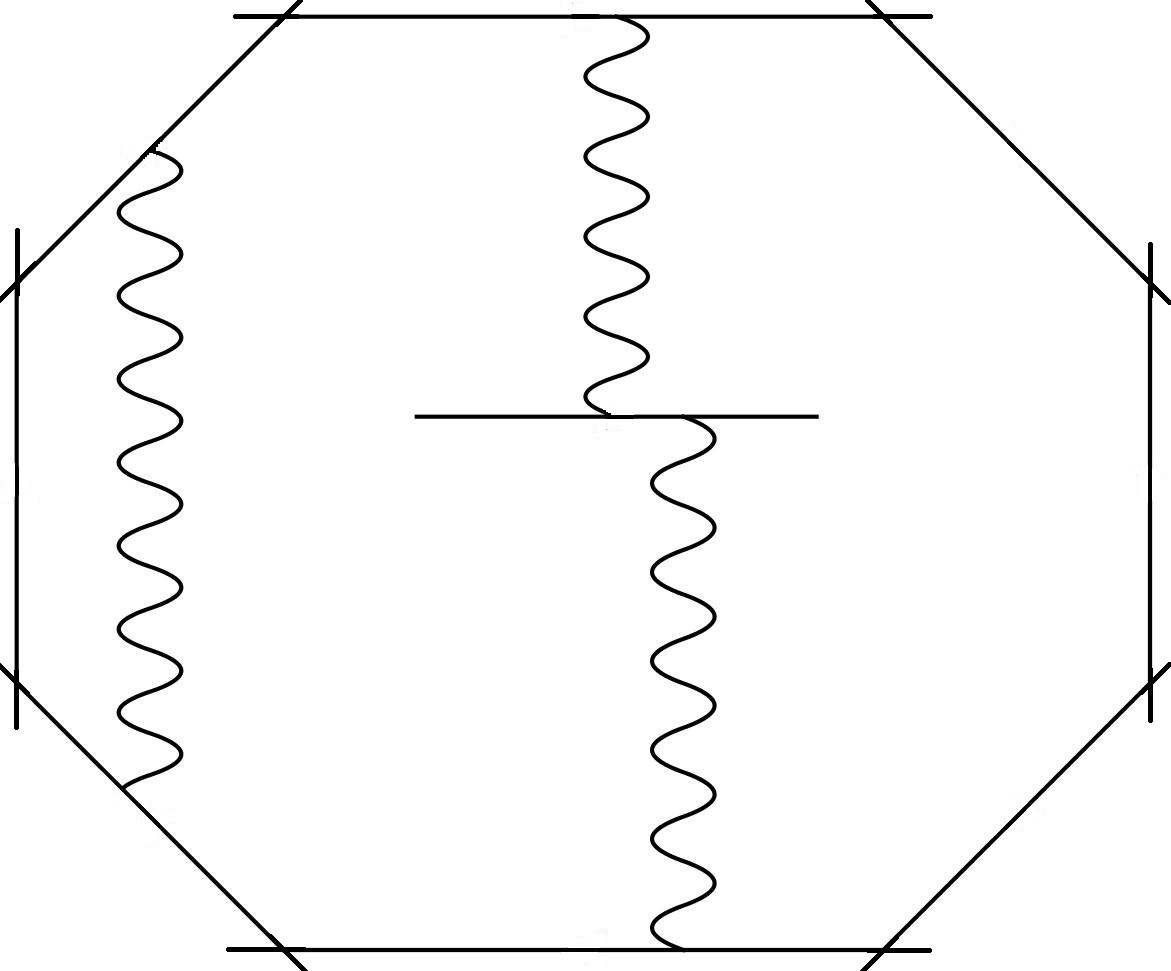}
\caption{ 1 loop NMHV 1st type.}
\label{nmhv1}
\end{center}
\end{figure}

\begin{figure}[h]
\begin{center}
\includegraphics[scale=0.23]{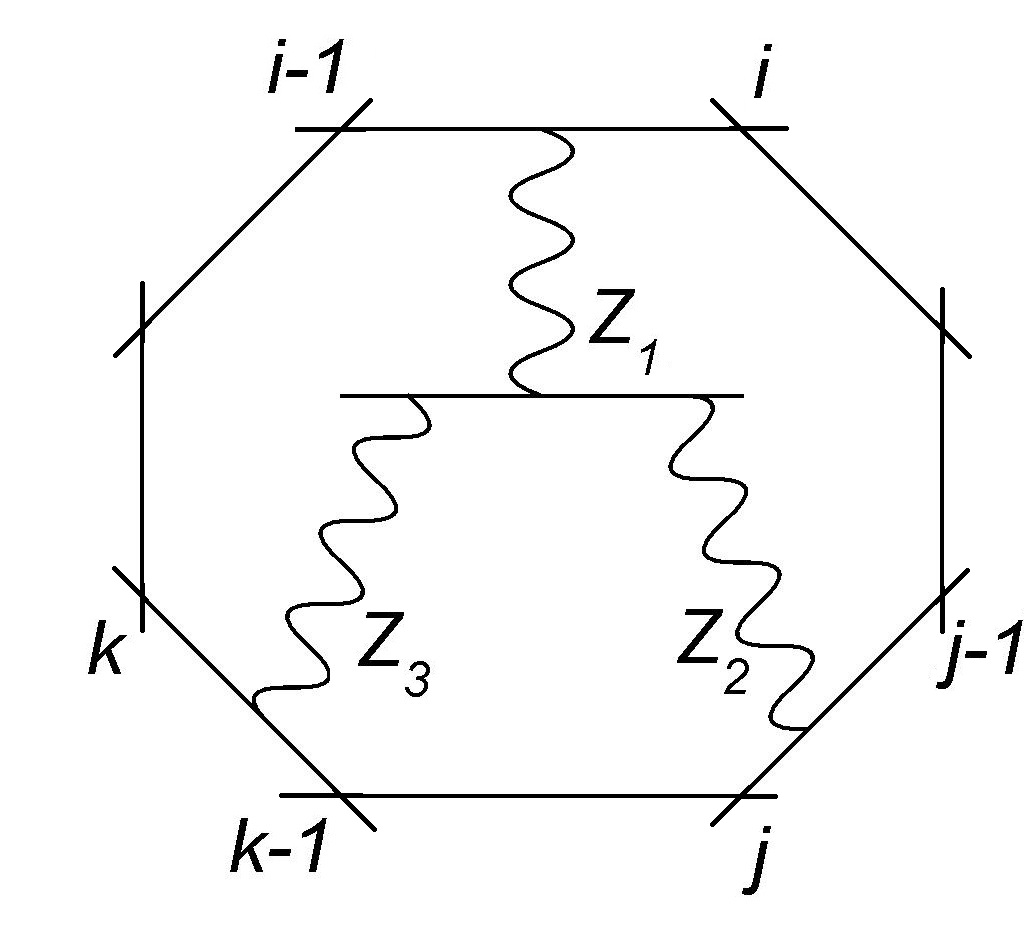}
\caption{ 1 loop NMHV 2nd type.}
\label{nmhv2}
\end{center}
\end{figure}

From our earlier arguments, at $L$ loops and MHV degree $k$ we will obtain an integrand with $4(L+k)$ dlog factors and $k$ $\bar \delta^{4|4}$'s.  So at NMHV we are left with one such $\bar\delta^{4|4}$.
At $L=1$ we will have one MHV vertex and three propagators.  There are two topological types of diagrams.  The first is where the additional propagator joins two sides and is  independent of  the MHV vertex (see figure \ref{nmhv1}), and the second is where there are three propagators attached to the MHV vertex (see figure \ref{nmhv2}).  

It is clear that for the first type we obtain a product of the 1-loop MHV and a tree level NMHV contribution.  The latter is four dlog integrals against a $\bar\delta^{4|4}$  and because the integrand and contours all factorize, the parameter integrals for the tree contribution can be performed separately to give an R-invariant as described in \eqref{Rinv} multiplied by the one loop MHV contribution. 

The only subtlety arises if the propagator corresponding to the R-invariant ends on the same egde as one of the other propagators. In this case, one of the parameters in the R-invariant integral will shift one of those in the 1-loop MHV integrals.  However, the integrals against the delta function will allow us to evaluate that parameter in terms of an invariant in the external parameters so that the shift will essentially be by a constant in the integrand yielding an integrand of the form
$$
\int\frac{\rd s_{1}}{s_{1}-a_1}\frac{\rd t_{1}}{t_{1}}\frac{\rd s_{2}}{s_{2}}\frac{\rd t_{2}}{t_{2}}
$$
where $a_1$ is a constant.

In the second type of diagram, we can immediately use \eqref{MHV-dlog} to find for the vertex with three propagators attached to twistors $Z_1$, $Z_2$ and $Z_3$
\be
\int _{\C^{5}}
\frac{\rd u_1\rd u_2}{u_1 u_2}
\frac{\rd s_{3}\rd t_{3}\rd u_3}{s_{3}t_3u_3}  
\bar{\delta}^{4|4}\left((1+t_3+s_3)Z_{*}+u_1Z_1+s_{3}u_2Z_2 +u_3 Z_3\right)\, .
\ee

These three twistors  then can then be integrated over sides $X_i$, $X_j$ and $X_k$ respectively to give
\begin{multline}
\int _{\C^{8}}
\frac{\rd u_1\rd u_2\rd u_3}{u_1 u_2u_3}
\frac{\rd s_1\rd s_2\rd s_{3}\rd t_2\rd t_{3}}{s_1s_2s_{3}t_2t_3}  \\
\bar{\delta}^{4|4}\left((1+t_3+s_3)Z_{*}+u_1(Z_{i-1}+s_1 Z_i)+s_{3}u_2(Z_{j-1}+s_2 Z_j) +u_3 (Z_{k-1}+t_2 Z_k)\right)\, .
\end{multline}
In fact the contour in this case is precisely the same as that for the 1-loop MHV after transforming to the  parameters 
$$
s_0=\frac1{u_1(1+s_1)}\, , \quad t_0=\frac 1{u_2(1+s_2)}\, , \quad s=s_1\, , \quad t=t_1\, ,
$$
with the remaining integrals being performed against the delta-functions.  It remains to be seen whether there is a canonical factorisation that allows us to factor off an R-invariant in some form for these diagrams.

\section{Conclusion} \label{conclusion}

In this paper, we have seen that the MHV loop amplitudes of planar $\mathcal{N}=4$ sYM can be expressed as integrals whose integrands have a dlog form.  In order to show this, we use the  holomorphic Wilson loop in momentum twistor space which is dual to the amplitudes. In particular, we find that the dlog form of the loop integrand follows directly from the Feynman rules for the twistor Wilson loop and the integration contour follows from reality constraints for the lines that carry the MHV vertices in momentum twistor space. We decribe this in detail for several examples including the one and two-loop MHV amplitudes and more briefly for the 1-loop NMHV amplitude. In the appendices, we discuss a toy model for the 1-loop MHV amplitude. We also demonstrate that our proposed expression in \eqref{kermit-fin} for the generic Wilson loop Feynman diagram contributing to the 1-loop MHV amplitude gives the correct symbol for the 1-loop MHV amplitude. 

Our analysis suggests that there might exist a general machine for performing the dlog integrals obtained from using the Feynman rules of the holomorphic Wilson loop, although we have not here gone so far as to produce one.  Two conceivable strategies are as follows.
One is to express the parameters that appear in the integrands as invariants involving the loop and external momenta, and to use Stokes theorem to perform integrations, reducing the integral essentially to combinatorics.   The second is to attempt to use the given parameters for the integration.  In this case, the external data of the loop amplitudes is encoded in the integration contours.  In both of these approaches one must confront the Feynman $i\epsilon$ prescription.  Although the simplest approach might be thought to be to simply use a Euclidean signature contour for the integration, it has thus far been easier to understand the Lorentz signature version.  In the Lorentzian approach, two parameters are naturally real for all the loop integrations to all loop orders and the task is to give the simplest specification of the $i\epsilon$ prescription for these parameters. In \cite{upcoming}, we will describe the $i\epsilon$ prescription in greater detail  and use it to derive \eqref{kermit-fin}. Since all MHV vertices have the same structure in our approach, a proper understanding of the MHV 1-loop example should suffice for a general understanding.

Another task is regularisation. In the 1-loop case for example, generic diagrams are finite but when the two propagators are attached to adjacent sides, the integral is divergent (but when the propagators are attached to the same side, the contribution can be taken to vanish  \cite{Brandhuber:2004yw}).  The simplest regularisation in our context will be mass regularisation \cite{Alday:2009zm} as providing an appropriate mass term should be not be significantly  harder than providing an $i\epsilon$.

For non-MHV amplitudes, we have shown that the integrand can be written in dlog form multiplied by delta functions but more work is required to see of the loop integrand can be further simplified. It appears that the dlog form of loop amplitudes can also be derived from the Grassmannian integral formula of $\mathcal{N}=4$ sYM \cite{ArkaniHamed:2012nw}. This suggests that there should be some direct way to relate the Grassmannian integral formula to the Wilson loop in twistor space.

\section*{Acknowledgments}

We are grateful to Nima Arkani-Hamed, Jake Bourjailly, Simon Caron-Huot, Dave Skinner and Jaroslav Trnka for freely sharing their ideas with us and to whom much of this material will be familiar. We would also like to thank Andreas Brandhuber, Gabriele Travaglini, and Congkao Wen for useful discussions. AL is supported by a Simons Postdoctoral Fellowship; LM is supported by a Leverhulme Fellowship and EPSRC grant number EP/J019518/1. 



\appendix

\section{A toy model for the evaluation of Kermit} \label{toy}

As a toy model for Kermit, let's evaluate the following integral\footnote{We thank Nima Arkani-Hamed for pointing out this toy model to us and for further discussions.}:
\be{}
\int_{S^2}  \rd \log s\,  \rd\log t \quad \mbox{over contour} \quad s=\frac{a\bar{t}+b}{c\bar{t}+d}
\ee
This can be done by expressing the integrand on the contour as the exterior derivative of
\be{}
\log t \,\rd  \log \left( \frac{a\bar{t}+b}{c\bar{t}+d}  \right) =\log t\left( \frac 1{\bar{t}+b/a} -\frac 1{\bar{t}+d/c}\right )\, \rd \bar t\, \, .
\label{integrand}
\ee
However, this form must have a cut from $t=0$ to $t=\infty$ and furthermore has poles at $\bar t=-b/a$  and $-d/c$.  So in order to use Stokes theorem, we must cut out an $\epsilon$-neighbourhood of the cut and the poles.  We can then use Stokes to reduce the integral to a contour integral around each pole and the cut.  The contour integral around the cut can then be reduced to a line integral along the cut as the contribution from the logarithmic singularities at the end vanishes as $\epsilon\rightarrow 0$, whereas the jump across the cut is $2\pi i$.  The contributions from the poles as $\epsilon\rightarrow 0$ is similarly given as $\mp 2\pi i\log t$ evaluated at the poles (noting the anti-holomorphic dependence on $t$) yielding
\be 
2\pi i\left(\int _0^\infty \rd \log \left( \frac{a\bar{t}+b}{c\bar{t}+d}  \right) - \log {\left(- \frac{\bar{b}}{\bar{a}} \right)} +\log {\left(-\frac{\bar{d}}{\bar{c}}\right)}\right) =2\pi i\ln\left|\frac{ad}{bc}\right|^2 \, .
\ee   

The toy model above resembles the $(s,t)$ integral in eq \ref{kermitfinal}. The reality condition in eq \ref{stconstraint} expresses $s$ as a mobius transform of $\bar t$, so the poles in $s$ and $t$ are complex and integrable and no prescription needs to be made to regulate these singularities in the integrand.  On the other hand, the integrals with respect to $s_0$ and $t_0$ have real poles and hence do need to be regulated. The poles in $s_0$ and $t_0$ are in fact associated with physical propagators and need to be regulated according to the Feynman prescription. Introducing such a prescription will lead to modifications of the contour described in eqs \ref{real0} and \ref{stconstraint}.  

\section{The one loop  symbol} \label{reverse}

The symbol of generic terms in the MHV amplitude is given by \cite{Gaiotto:2011dt,Bullimore:2011kg, CaronHuot:2011ky}
$$
d R_n= \sum_{ij} \log \left( \frac{x_{i+1\,j}^2 x^2_{j+1\,i}}{x_{i+1j+1}^2x^2_{i\, j}}\right)\rd \log (i-1\, i \, i+1\,j) \,
$$
Up to normalization, $\left(i-1ii+1j\right)=a_{ji}$ so the symbol can be written as

\begin{eqnarray*}
d R_n &=&\sum_{ij}\left[\left(\log x_{i+1j}^{2}+\log x_{j+1i}^{2}-\log x_{i+1j+1}^{2}-\log x_{ij}^{2}\right)d\log a_{ji}\right]\\
&=& \sum_{ij}\log x_{ij}^{2}d\log\frac{a_{ji-1}a_{j-1i}}{a_{ji}a_{j-1i-1}}
\end{eqnarray*}
where we relabeled dummy indices to obtain the second equality. Since
Kermit only knows about the lines $X_{i}$ and $X_{j}$, each term in
the sum above should correspond to the part of the symbol that Kermit must
provide. Furthermore, since

\[
x_{ij}^{2}=\frac{\left(ii-1jj-1\right)}{\left\langle ii-1\right\rangle \left\langle jj-1\right\rangle },\]
we see that $\log x_{ij}^{2}=\log\left(ii-1jj-1\right)+$ terms
which break dual conformal symmetry and therefore cancel out of the
sum since $R_{n}$ is a dual conformal invariant. Hence, \begin{equation}
dR_{n}=\sum_{ij}\log\left(ii-1jj-1\right)d\log\frac{a_{ji-1}a_{j-1i}}{a_{ji}a_{j-1i-1}}.\label{eq:symbol}\end{equation}
To better conform with our notation, note that using reality we can
write 

\[
\left(ii-1jj-1\right)=a_{j-1i}a_{ji-1}-a_{ji}a_{j-1i-1}\]
so eq \ref{eq:symbol} reduces to

\begin{equation}
dR_{n}=\sum_{ij}\log\left(u_{ij}-v_{ij}\right)d\log\frac{u_{ij}}{v_{ij}}\label{eq:symbuv}\end{equation}
where we have defined the variables
\[
u_{ij}=a_{ij-1}a_{i-1j},\,\,\, v_{ij}=a_{ij}a_{i-1j-1}.\]
Equation \eqref{eq:symbuv} implies that the symbol of Kermit is given by 
\begin{equation}
\left(u_{ij}-v_{ij}\right)\otimes\frac{u_{ij}}{v_{ij}}+\mbox{c.c.}
\label{symbol-fin}
\end{equation}
For a brief review of symbols, see for example section 2.1 of \cite{Gaiotto:2011dt}.  

We will now verify that the expression for Kermit in \eqref{kermit-fin} leads to the correct symbol for the 1-loop MHV amplitude. For convenience, we reproduce \eqref{kermit-fin} below: 
\begin{equation}
\Li_{2}\left(\frac{ia_{ij-1}}{v_{*}}\right)+\Li_{2}\left(\frac{ia_{i-1j}}{v_{*}}\right)-\Li_{2}\left(\frac{ia_{ij}}{v_{*}}\right)-\Li_{2}\left(\frac{ia_{i-1j-1}}{v_{*}}\right) + \mbox{c.c.}\, 
\label{kermit-fin2}
\end{equation}
where
\[
v_{*}=i\frac{u_{ij}-v_{ij}}{a_{*}},\,\,\,a_{*}=a_{i-1j}+a_{ij-1}-a_{ij}-a_{i-1j-1}.
\] 
Let's compute the symbol of the third term in \eqref{kermit-fin2}. The symbols of the other terms will be similar. The symbol for this term is given by  
\begin{equation}
\left(1-\frac{ia_{ij}}{v_{*}}\right)\otimes\frac{ia_{ij}}{v_{*}}=\left(u_{ij}-v_{ij}-a_{ij}a_{*}\right)\otimes\frac{a_{ij}a_{*}}{u_{ij}-v_{ij}}-\left(u_{ij}-v_{ij}\right)\otimes\frac{a_{ij}a_{*}}{u_{ij}-v_{ij}}.
\label{Li2}
\end{equation}
Note that
\[
u_{ij}-v_{ij}-a_{ij}a_{*}=\left(a_{ij}-a_{ij-1}\right)\left(a_{ij}-a_{i-1j}\right)
\]
so the first term on the right-hand-side of \eqref{Li2} can be written as 
\[
\left(a_{ij}-a_{ij-1}\right)\otimes\frac{a_{ij}a_{*}}{u_{ij}-v_{ij}}+\left(a_{ij}-a_{i-1j}\right)\otimes\frac{a_{ij}a_{*}}{u_{ij}-v_{ij}}.
\]
It is not difficult to see that these terms will be cancelled by corresponding terms arising from the other three dilogs in \eqref{kermit-fin2} when we sum over $i,j$. Furthermore, the second term in \eqref{Li2} can be written as 
\[
-\left(u_{ij}-v_{ij}\right)\otimes a_{ij}-\left(u_{ij}-v_{ij}\right)\otimes\frac{a_{*}}{u_{ij}-v_{ij}}.
\]
After summing over the four dilogs in \eqref{kermit-fin2}, the first term in the equation above leads to the symbol in \eqref{symbol-fin} and second term in cancels out. Hence, our conjecture for Kermit in \eqref{kermit-fin} indeed gives the correct symbol for the 1-loop MHV amplitude. We will obtain \eqref{kermit-fin} from a direct calculation in \cite{upcoming}.

\end{document}